\documentclass[12pt,a4paper]{article}
\usepackage{amsmath}
\usepackage[font={footnotesize,it}]{caption}
\usepackage[top=1in, bottom=1in, left=1in, right=1in]{geometry}
\DeclareCaptionStyle{italic}[justification=centering]{labelfont={bf},textfont={it},labelsep=colon}
\usepackage[title]{appendix}

\usepackage{multirow, fancyvrb}
\usepackage{tabularx}
\usepackage[table]{xcolor}
\usepackage{pdflscape,microtype}

\usepackage{colortbl}
\usepackage{graphicx,psfrag,epsf,textcomp}
\usepackage{threeparttable}

\usepackage{adjustbox}

\usepackage{lineno}
\usepackage{setspace}
\usepackage{enumerate}
\usepackage[round]{natbib}
\usepackage{url,xcolor} 
\usepackage{booktabs, bm, paralist,mathpazo,tikz,longtable,microtype}
\usepackage{xcolor,soul,amsfonts,setspace,mathrsfs,mathtools,placeins}
\usepackage{multirow,array}
\usepackage{float}
\usepackage{subcaption}
\usepackage{longtable}

\usepackage[pdftex,colorlinks=true]{hyperref}
\definecolor{darkblue}{rgb}{0,0,.6}
\definecolor{a0}{rgb}{0.0, 0.5, 0.0}
\definecolor{bistre}{rgb}{0.24, 0.17, 0.12}
\definecolor{amethyst}{rgb}{0.6, 0.4, 0.8}
\definecolor{Rcolor}{RGB}{150,160,190}
\definecolor{blush}{rgb}{0.87, 0.36, 0.51}
\definecolor{brightturquoise}{rgb}{0.03, 0.91, 0.87}
\definecolor{burntorange}{rgb}{0.8, 0.33, 0.0}
\definecolor{codashade}{RGB}{230,235,250}
\hypersetup{citecolor=darkblue,linkcolor=darkblue,urlcolor=darkblue}
\newcommand*\patchAmsMathEnvironmentForLineno[1]{%
  \expandafter\let\csname old#1\expandafter\endcsname\csname #1\endcsname
  \expandafter\let\csname oldend#1\expandafter\endcsname\csname end#1\endcsname
  \renewenvironment{#1}%
     {\linenomath\csname old#1\endcsname}%
     {\csname oldend#1\endcsname\endlinenomath}}%
\newcommand*\patchBothAmsMathEnvironmentsForLineno[1]{%
  \patchAmsMathEnvironmentForLineno{#1}%
  \patchAmsMathEnvironmentForLineno{#1*}}%
\newcommand{\Rlogo}{\protect\includegraphics[height=1.8ex,keepaspectratio]{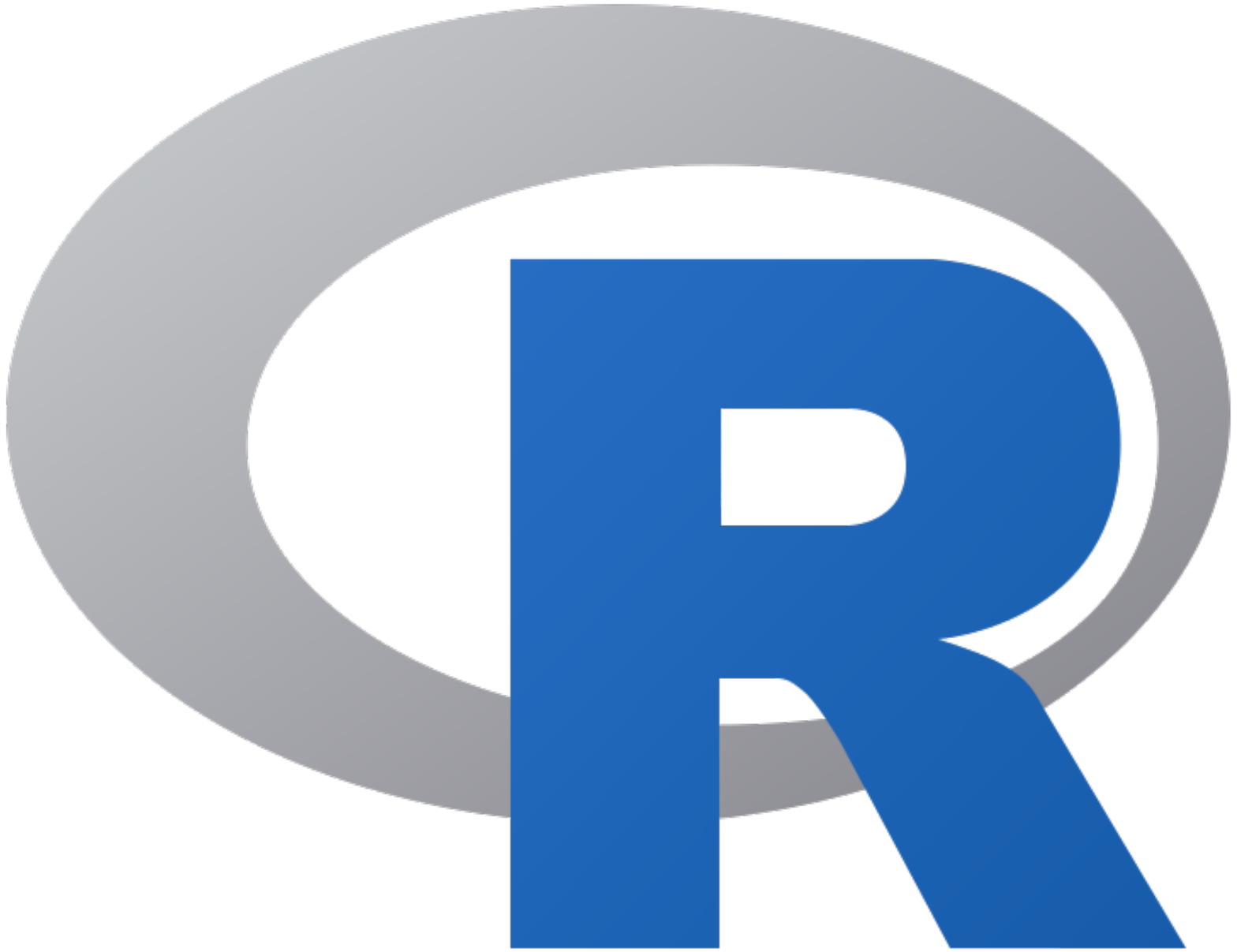}}

\AtBeginDocument{%
\patchBothAmsMathEnvironmentsForLineno{equation}%
\patchBothAmsMathEnvironmentsForLineno{align}
\patchBothAmsMathEnvironmentsForLineno{flalign}%
\patchBothAmsMathEnvironmentsForLineno{alignat}%
\patchBothAmsMathEnvironmentsForLineno{gather}%
\patchBothAmsMathEnvironmentsForLineno{multline}%
}

\newcommand{\blind}{0}

\graphicspath{{images/}}

\newcolumntype{L}[1]{>{\raggedright\let\newline\\\arraybackslash\hspace{0pt}}m{#1}}
\newcolumntype{C}[1]{>{\centering\let\newline\\\arraybackslash\hspace{0pt}}m{#1}}
\newcolumntype{R}[1]{>{\raggedleft\let\newline\\\arraybackslash\hspace{0pt}}m{#1}}

\newsavebox\CBox

\date{}
\usepackage{orcidlink}

\begin{document}

\def\spacingset#1{\renewcommand{\baselinestretch}
{#1}\small\normalsize} \spacingset{1}

\if0\blind
{
\title{\bf Climate sensitivity analysis -- A case study from forty years of US compositional cause of death data}
\author{\normalsize Michelle Dong\qquad Aaron Bruhn\qquad Francis Hui \\
\normalsize Research School of Finance, Actuarial Studies and Statistics \\
\normalsize Australian National University \\
\\
\normalsize Han Lin Shang \\
\normalsize Department of Actuarial Studies and Business Analytics \\
\normalsize Macquarie University
}
\date{\normalsize \today}
\maketitle
}\fi

\if1\blind
{
\title{\bf Climate sensitivity analysis -- A case study from forty years of US compositional cause of death data}
\author{}
\date{}
\maketitle
} \fi

\begin{abstract}{
Emerging climate risks pose pressing challenges for insurers, governments, and businesses worldwide, as they face growing uncertainty in quantifying the impacts of climate change from physical risks. This paper aims to understand the impact of specific climate factors on mortality by cause for subgroups of the United States (US) population, through sensitivity and scenario analysis based on an increase in temperature and sea level extremes. We apply compositional data analysis (CODA) techniques to examine cause-specific deaths, treating the density of deaths as a set of dependent, non-negative values that sum to one. We couple CODA with principal component analysis on the climate factors as a means of dimension reduction, and fit generalised additive models to better reflect the non-linear relationships between the dimension-reduced principal scores and mortality by cause. The results of our analysis indicate climate-related factors have varying impacts by cause and ages within each cause, with more pronounced increases in the proportions of deaths from hypertensive heart disease as temperature and sea level extremes increase. Scenario analysis also indicates that an increase in temperature high extremes, sea level, and rainfall, in conjunction with a decrease in low temperature extremes, lead to offsetting impacts on the proportion of deaths between climate-related causes, but less offsetting by age within causes. Furthermore, the impacts on the proportions of death are more pronounced for ages between 55 and 95, reinforcing the observation that climate-related risks have a greater impact on older (and potentially more vulnerable) subgroups of the population. For life insurers specifically, these results are consistent with the natural hedge that arises between annuity and protection products, in light of increasing climate risk. 
}
\vspace{.1in}

\noindent \textit{Keywords}: compositional data analysis; cause of death; climate scenario modelling; climate-related risks; mortality modelling and forecasting
\end{abstract}

\spacingset{1.5}
\newpage

\section{Introduction}\label{sec:introduction}

Understanding climate-related risks is an increasingly relevant and topical field for governments, academics, and industry. Life insurance is no different, and faces the same challenges when it comes to understanding and managing climate-related risks and impacts. Given the systemic and unprecedented nature of these changes, traditional methods of analysis that heavily rely on extrapolating general trends from historical data are becoming less informative. In its place, sensitivity and scenario analysis has increasingly become a pivotal method to understand the impact of climate-related risks on mortality and morbidity, enabling researchers and users to consider a range of likely outcomes and to make informed decisions based on the range of potential outcomes. In constructing counterfactual scenarios, it is imperative to leverage known relationships between climate risk factors and climate responses, as well as to take into account the variability around these relationships \citep{BOE25}.

There exists extensive research in understanding the impact of climate-related risks on life and health, ranging from national or global scale studies on climatic factors and impacts on GDP, to risk-specific studies, e.g., the impact of changes in the temperature range on mortality by age; see, for instance \citet{WB25}, \cite{LKS+20}. 
However, an area that has not yet garnered extensive quantitative exploration is quantifying the impact of specific climate factors, e.g., temperature and precipitation, on different causes of death across age bands. This is critical for elucidating climate impacts across differing areas of exposure for insurers, and identifying likely trends in health impacts for specific subgroups.

In this article, we investigate climate risk factors and their implications on mortality by cause for subgroups of the United States (US) population, and apply scenario and sensitivity analysis to this. We evaluate mortality impact for males and females separately, using death counts by cause as the response, and including various climate indices such as the temperature and precipitation factors underlying the Actuaries Climate Index as the explanatory variables of interest. We explicitly account for the compositional constraint inherent in the death count by cause responses. That is, the elements of the vector of counts sum up to a constant value, meaning they only contain relative information and the avoided deaths of one cause lead to the increase of deaths from other causes \citep{A86}. To handle this, we apply compositional data analysis (CODA) techniques to model the relationship between climate factors and causes of death, based on the transformation of compositional data from a constrained space to an unconstrained space 
\citep{TPW11,G21}. Two models within the CODA framework are investigated, namely a linear model and a generalised additive model \citep[GAM,][]{mgcv,harezlak2018semiparametric}, with the latter, more flexible approach used downstream for scenario analysis on a climate counterfactual to assess the impact of deaths for different subgroups for the US.

\subsection{Literature review} \label{subsec:litreview}
 
The literature on understanding climate risks has grown over the past decade, particularly in recent years as governments and regulators across the globe have worked to develop climate risk assessments and climate scenario testing requirements to better understand individuals' and companies' exposures to climate risks. Below, we review several of the most prominent ones relevant to the actuarial science field.

In 2023, the Swiss Re Institute investigated the impact of climate change on mortality through a mix of quantitative and qualitative research. It sets the expectation for health and mortality impacts of climate change to ``play out gradually with an incremental impact on life and health risks", and identified the largest climate drivers on life and health as extreme heat, air pollution, and increased exposure to infectious diseases spread by non-human vectors \citep{SRI23}. The resulting paper identified the key causes likely to be affected by climate-related risks as cardiovascular and respiratory, and driven by temperature and rainfall extremes. Furthermore, they noted that the trajectories of air pollution and infectious diseases are ``much more uncertain", relative to extreme heat. At the same time, they acknowledged that attempts to quantify mortality impacts can be speculative and subject to wide uncertainty intervals, with the focus at the time of writing likely to be on long-term trends rather than shock events.

The findings of the Swiss Re Institute paper are corroborated by other empirical research. For instance, in terms of temperature impacts, \cite{ZGY+21} investigated excess mortality caused by global variations in temperature, and found that non-optimal temperatures are associated with a substantial mortality burden. In a similar vein, the US Environmental Protection Agency recently performed an analysis based on data collected for heat-related deaths, with results showing that such deaths reached new highs in 2021 and 2022, as reflected in cardiovascular causes and other factors i.e., Lyme disease and West Nile virus, related to vector-borne diseases \citep{EPA25}. Going further back, quantitative research published by the World Health Organisation (WHO) in 2014 suggested a significant impact of climate change on mortality \citep{WHO14}. The assessment was performed by estimating future cause-specific mortality for 2030 and 2050, with and without climate-scenario overlays. Climate scenario overlays were then added to reflect heat-related mortality among the elderly, coastal flooding, diarrhoeal disease, malaria, dengue, and malnutrition. These scenarios were based on A1b from the Special Report on Emission Scenarios, rather than the later IPCC scenarios which were not available at the time. 

More recent findings from the WHO also indicate that temperature and precipitation changes enhance the spread of vector-borne diseases. These are expected to be more severe moving forward without preventive actions. Additionally, other risks exacerbated by climate change include mental health, food insecurity, and costal flooding. The WHO also notes that modelling challenges persist, especially in capturing emerging risks such as drought and migration pressures. Related research from the WHO further attributes close to 40\% of heat-related deaths to human-induced climate change, with the percentage observed to increase particularly among those over age~65 \citep{WHO23}. Elsewhere, \citet{HBW+11} reviewed projections of future heat-related mortality under climate change scenarios, and concluded that scenario-based projection research is expected to contribute to understanding and managing the potential impacts of climate change on heat-related mortality. \citet{GGS+17} assessed projections of temperature-related excess mortality under climate change scenarios, and found that on average, there is expected to be a net increase in temperature-related excess mortality under high-emission scenarios, but with important geographical differences disproportionately affecting warmer and poorer regions of the world.

A particularly recent and relevant work comes from Australia's first National Climate Risk Assessment \citep[NCRA,][]{ACS25}. This is the first report to deliver a synthesised understanding of the climate risks Australia is experiencing, and provides insights into how climate risk affects different sectors and regions of the country. In particular, the key systems that the assessment considers include defence and national security, economy, trade and finance, and health and social support, among others.
Australia's NCRA identifies current health and social support risks as moderate, but expects these to increase to severe by 2050. 

In relation to insurers and insured risks, the Prudential Regulation Authority (PRA) in the UK issued a Policy Statement (PS25/25) following consultation in 2025, along with Supervisory Statement (SS5/25). The latter is intended to enhance how banks and insurers manage and build resilience against climate-related risks, with one notable requirement being to advance the use of climate scenario and sensitivity analysis, enabling firms to demonstrate an understanding of how climate scenario outputs inform business decisions \citep{PRA25, PRA25b}. Like the Australian NCRA report above, the PRA Policy Statement also acknowledges that impacts of climate change are expected to grow over time, and affect banks and insurers through direct losses and business model changes. For life insurers in particular, the most substantial threat is that liabilities may arise from increased morbidity and mortality from heat waves, along with other indirect impacts on mortality and morbidity from rising temperatures, such as an increase in vector-borne diseases. 

As two contemporary sources, both Australia's first NCRA report and the UK PRA's PS25/25 and SS5/25 statements reflect the growing motivation to better understand climate risk impacts for life insurers through scenario analysis. Moreover, the importance of the role of actuaries in understanding and managing climate risk is growing, with early seminal articles published by the International Association of Actuaries (IAA) stressing the need for actuaries to play a role in identifying, managing, measuring, and reporting climate-related risks \citep{IAA00}. Subsequent papers released by the IAA provide insights into the requirements and tools available to enable actuaries to perform meaningful climate-related sensitivity and scenario analysis \citep{IAA21, IAA22}. This article contributes to this strand of literature, by considering climate scenario analysis in the context of understanding causes of death through a CODA-based approach. 

Turning to the statistical methods underlying scenario and sensitivity analysis, historically many approaches to modelling and forecasting mortality do not explicitly account for dependencies between causes of death, i.e., they treat cause-specific life-table deaths as non-compositional in nature \citep{KEK+19}. A main drawback of this approach when used for forecasting is that subcomponent forecasts may diverge, due to the inability of subcomponent models to forecast consistently with one another \citep{BCO17}. To circumvent this issue, CODA techniques have emerged over the past decade in the actuarial science literature to better account for the explicit compositional structure of cause of death mortality data.
\citet{KEK+19} used CODA approaches to forecast causes of death from cancer, introducing models to forecast deaths by cause over time with age-, period-, and cause-specific weights in their application. They found that the best-performing CODA model produced more accurate forecasts than non-CODA approaches such as the classic Lee-Carter model. \citet{KEB+20} applied CODA methods for longevity forecasting where the composition was split into subgroups by socioeconomic groups. They again found CODA models consistently produced forecasts with lower errors than non-CODA approaches. More recently, \citet{BSS+22} modelled healthy life expectancy using a combination of Sullivan's method plus CODA, providing insights for actuaries and other professionals in life insurance with forecasts informing total life expectancy, split into disability-free and severe disability-free states. In a similar vein, \citet{DSH+25} applied CODA techniques combined with the $\alpha$-transformation \citep{TPW11} to model deaths by cause when there are zero counts in subgroups, providing an improved approach to handling zeros in compositional data.

Finally, in relation to the current literature on climate scenario analysis, 
the seminal works of \citet{CISL22} and \citet{IAA21} elaborate on approaches to defining climate scenarios, with reference to available data, modelling techniques, and acknowledging the likely need for actuaries to work with other professions in developing and considering climate scenarios. Other applications include \citet{BBC+24}, which takes a data science approach to climate change risk assessment, by considering the impact on (re)insured losses from flood risk due to climate change. 
More recently, \citet{AAA25} provide insight into how practitioners could consider climate scenario analysis for US Own Risk Solvency Assessment. This paper makes reference to the range of available scenarios for actuaries -- from the IPCC scenarios, which do not assign probabilities but rather refer to broad narratives that describe the probability of physical outcomes, to more bespoke and specific scenarios developed by regulatory bodies for the financial services sector, including the Bank of England's Climate Biennial Exploratory Scenario (CBES) for the largest UK banks and insurers, and Canada's Office of Superintendent of Financial Institutions Standardised Climate Scenario Exercise. It highlights the range of available scenarios for actuaries, and emphasises the need to tailor and adjust scenarios. 
At the same time, there has also been some critique of using existing scenarios. Notably, \citet{IFOA23} highlights some challenges with modelling, including how existing scenarios may underestimate climate risk and introduce the risk of group think. The paper also suggests the importance of understanding assumptions underpinning the models and limitations, and emphasises the growing concern that ``time is too short to wait for models that are perfect".

\subsection{Main findings and contributions}

With the above literature review in mind, the main contributions and findings of this article can be summarised as follows: 
\begin{asparaenum}[1)]
\item We apply CODA regression methods, in particular coupling CODA techniques with principal components analysis (PCA) for dimension reduction and GAMs for non-linear modelling, 
to model deaths by cause and identify differences in the mortality response by cause and age. This statistical approach forms the basis for an alternative approach to considering climate scenarios, stress-testing mortality, and understanding impacts by cause.
\item Our sensitivity and scenario analysis capture the varying impacts of a range of climate factors for multiple causes of death and age bands within causes across the US population, and validate a number of qualitative conclusions in the existing literature around the impact of climate risks for specific subgroups of the population.
\item Scenarios utilising stress tests on temperature and sea levels suggest that, with more extreme temperatures and rainfall, there will likely be increases in the proportion of deaths for hypertensive heart disease and Chronic Obstructive Pulmonary Disease (COPD) in older ages, with possible offsetting deaths for other cardiovascular-related causes.
\item The impacts of more extreme climate factors are more pronounced among adults aged 65 and older, a feature observed across all climate-related causes identified. This observation reinforces the literature that climate-related risks have a greater impact on older, potentially more vulnerable subgroups of the US population. 
\end{asparaenum}

The remainder of this article is structured as follows:  Section~\ref{sec:data} introduces the various US data sources used for the application, including the compositional response, explanatory factors, and results from applying PCA to the latter. Section~\ref{sec:methodology} details the CODA framework and its combination with linear models and GAMs, while Section~\ref{sec:application} applies this to the US mortality by cause data and discusses the main results of the sensitivity analysis. Finally, Section~\ref{sec:discussion} offers some concluding remarks and avenues for future research.

\section{Data}\label{sec:data} 

In this section, we provide details of our response, i.e., compositional cause of death data (Section \ref{subsec:US_data}), along with the explanatory climate and macroeconomic factors we included in CODA regression models for sensitivity and scenario analysis (Section \ref{subsec:explanatorydata}). 

\subsection{US cause of death response}\label{subsec:US_data}

We sourced male and female US cause of death mortality data from the Human Cause of Death Data series \citep[HCD,][]{HMD24}, which contains over 40 years of historical cause of death count data for the United States from 1979 to 2023. For the purposes of this article, we set the primary causes of death of interest to cardiovascular, respiratory, and vector-borne diseases, given the known links of these to climate-related risks as reviewed in Section~\ref{subsec:litreview}, while all other causes were grouped together into a separate ``Other" category. The cause of death grouping at the HCD's Intermediate cause level provides a suitable balance between data granularity and the inter  pretation of the results. In particular, Table~\ref{tab:hcd_mapping} shows the mapping between the selected climate-related cause identified above, and the ICD-10 classification \citep{ICD10}. Similar to \citet{DSH+25}, we grouped the ages into ten-year bands from 25 to 95. 
\begin{table}[htbp]
\centering
\tabcolsep 0.12in
\begin{tabularx}{\textwidth}{@{}>{\raggedright\arraybackslash}p{3.5cm}>{\raggedright\arraybackslash}X>{\raggedright\arraybackslash}p{5cm}@{}}
\toprule[1.5pt]
\textbf{Group} & \textbf{Intermediate HCD Cause} & \textbf{ICD-10 Mapping} \\
\cmidrule{2-3}
\multirow{4}{=}{Cardiovascular} 
& I031 Rheumatic heart diseases & I00–I09 \\
& I032 Hypertensive disease     & I10–I13 \\
& I033 Ischaemic heart diseases & I20–I25 \\
& I034 Other heart diseases     & I26–I51 \\\\
\multirow{4}{=}{Respiratory} 
& I039 Pneumonia                & J12–J18 \\
& I040 Other acute respiratory infections    & J00–J06, J20–J22, U04 \\
& I041 Other chronic obstructive pulmonary disease (COPD) & J40–J47 \\
& I042 Other diseases of the respiratory system & J30–J39, J60–J98 \\\\
\multirow{1}{=}{Vector-borne diseases}
& I006 Other and unspecified infectious and parasitic diseases & A20–A39, A42–A99, B00–B09, B25–B89, B91–B99 \\
\bottomrule[1.5pt]
\end{tabularx}
\caption{HCD Intermediate cause level and associated ICD-10 classification mappings. The three causes (cardiovascular, respiratory, and vector-borne diseases) were selected based on their identification with climate-related drivers in the existing literature. All other causes are grouped together as ``Other", including COVID-19.}
\label{tab:hcd_mapping}
\end{table}

\begin{figure}[htb]
\centering
\begin{subfigure}[t]{0.48\textwidth}
\centering
\includegraphics[width=\linewidth]
{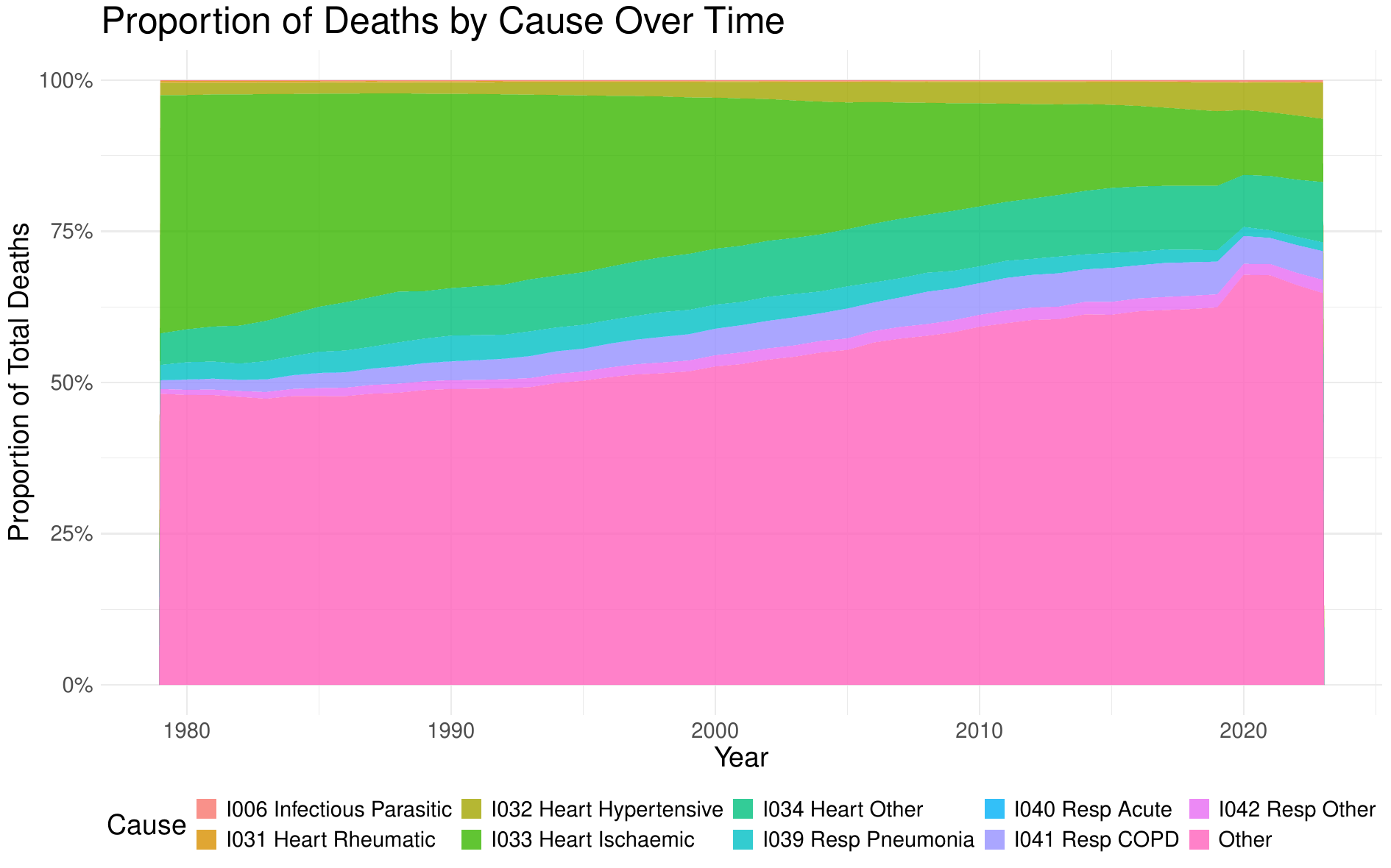}
\caption{Female deaths by cause (all ages)}
\end{subfigure}
\hfill
\begin{subfigure}[t]{0.48\textwidth}
\centering
\includegraphics[width=\linewidth]
{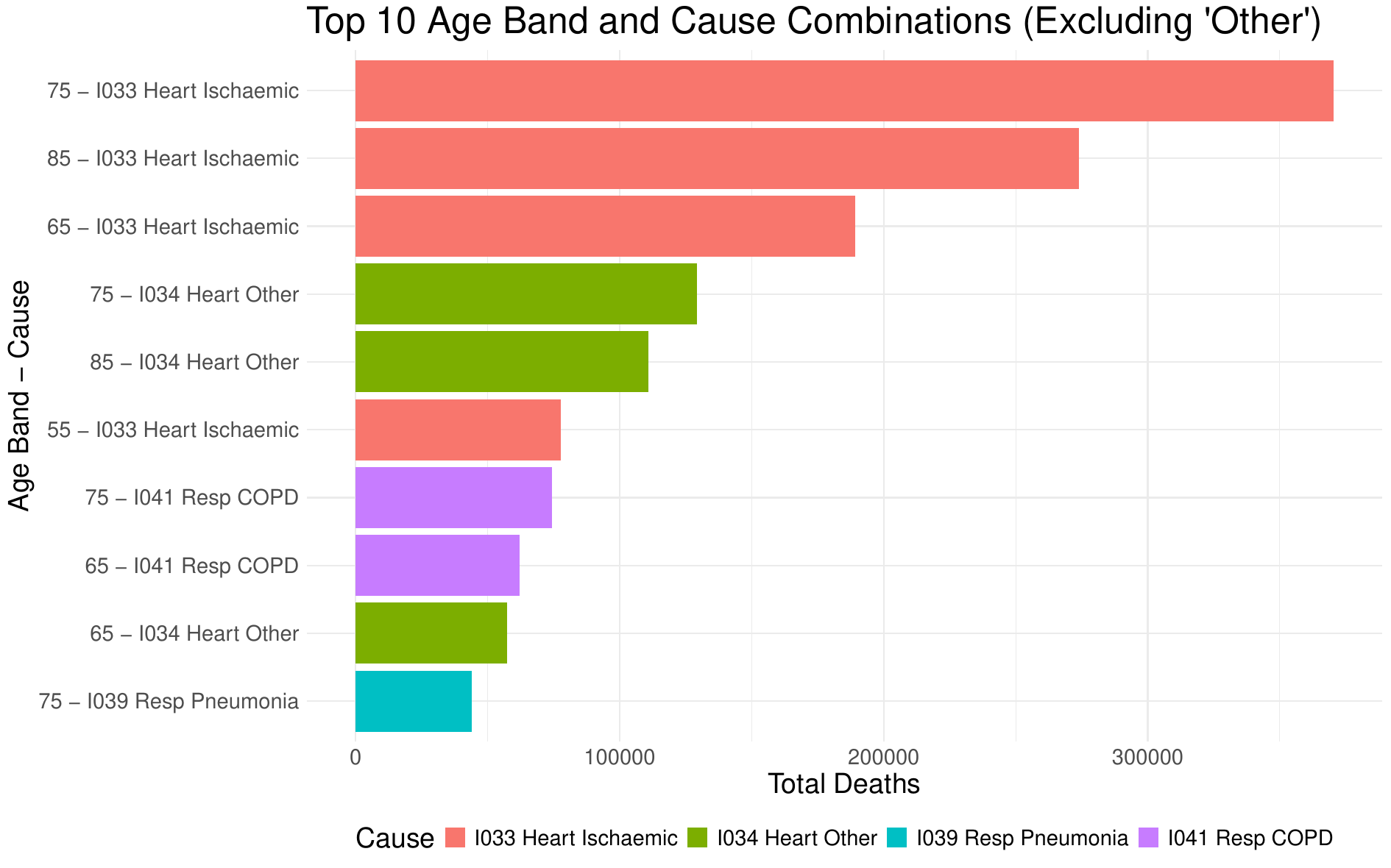}
\caption{Top ten causes and age bands (females)}
\end{subfigure}

\vspace{1em}
\begin{subfigure}[t]{0.48\textwidth}
\centering
\includegraphics[width=\linewidth]
{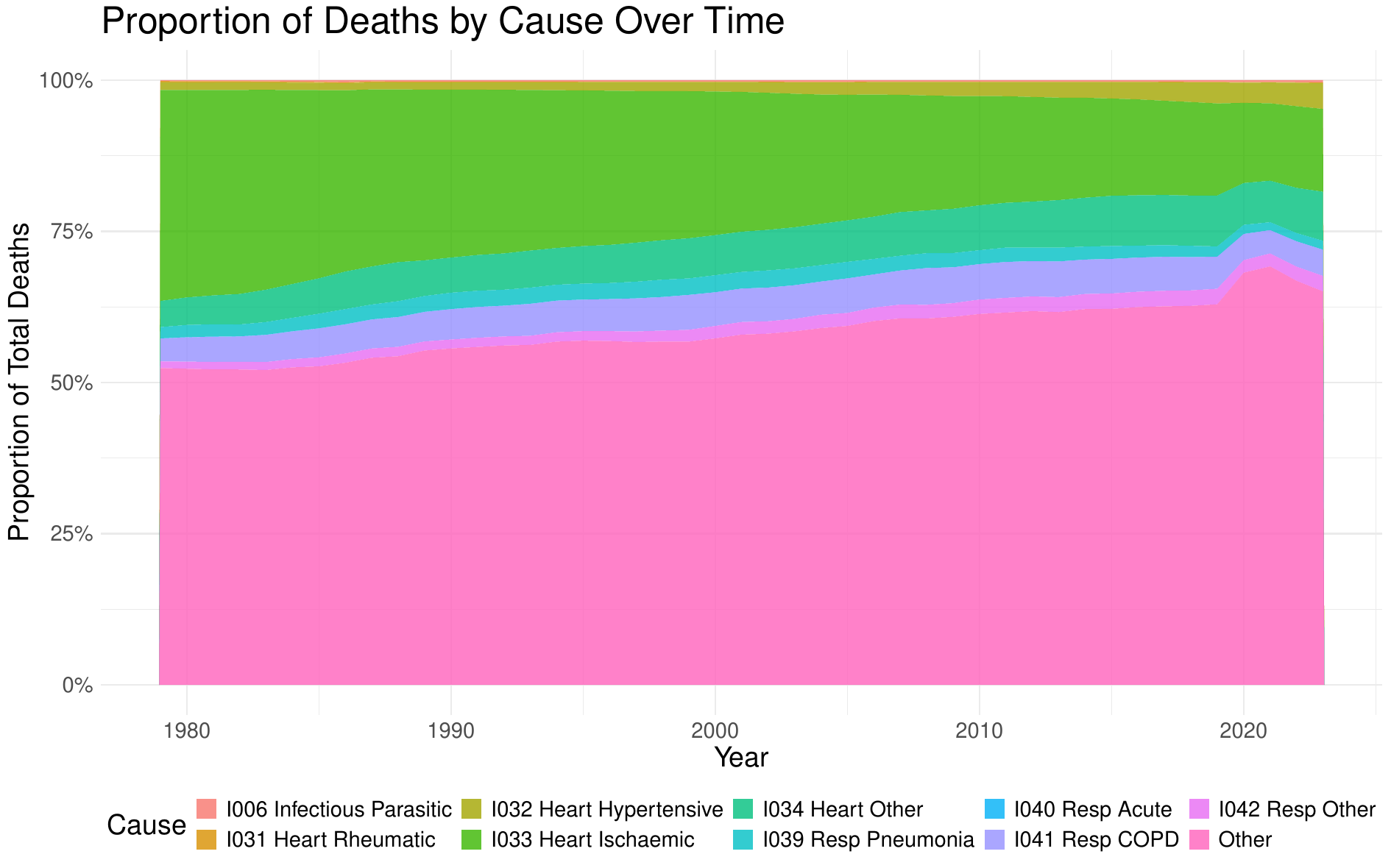}
\caption{Male deaths by cause (all ages)}
\end{subfigure}
\hfill
\begin{subfigure}[t]{0.48\textwidth}
\centering
\includegraphics[width=\linewidth]
{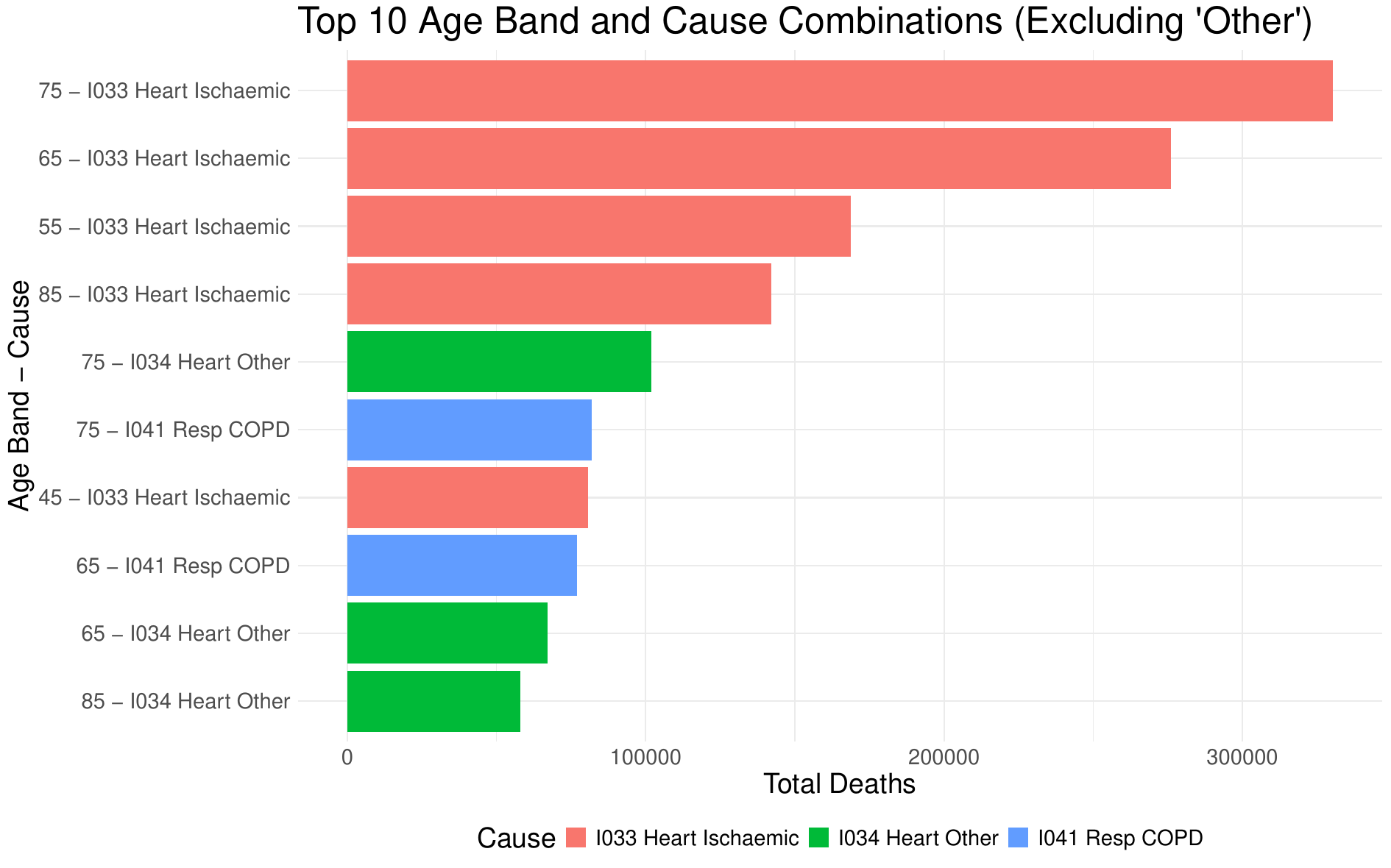}
\caption{Top ten causes and age bands (males)}
\end{subfigure}
\caption{Proportion of deaths by cause for the US population from 1979 to 2023. The charts on the left (panels a and c) present death proportions by cause, aggregated over all ages. The charts on the right (panels b and d) present the top ten groups in the composition by age band and cause, and provide insight into the largest cause and age combinations within the selected causes.}
\label{fig:compositional_data}
\end{figure}

Figure~\ref{fig:compositional_data} presents the aggregated death proportions for all ages from 1979 to 2023 for the US, together with the top ten groups in the composition for males and females (where a group is defined as a cause and age band combination). As observed, the number of deaths for some causes was small, even when aggregated across all ages. Also, the impact of the COVID years was apparent when examining at all-age cause of death charts (panels a and c). Regardless, the main causes of death were similar between males and females, with Ischaemic heart disease dominating the main causes, closely followed by other cardiovascular causes and COPD (panels b and d). Within the top ten age bands and cause groups, most age bands were over 65, except for Ischaemic heart disease, where ages 45 and 55 appeared for males and females, respectively. In terms of differences between males and females, pneumonia in the age 75 band ranked as the 10th most common cause among females, but not among males. 

\subsection{Explanatory variables}\label{subsec:explanatorydata} 

We sourced climate variables for the US from the Actuaries Climate Index \citep[ACI,][]{ACI25}, while macroeconomic factors from the World Bank database \citep{WB25} were extracted to capture wider non-climate-related trends. Although ACI factors were available monthly or seasonally, for this analysis we selected smoothed winter indices for the US region, in line with the location and frequency of cause of death data. Alternatives to the smoothed winter index were tested, though the results remained broadly similar when annual data were considered. More importantly, we considered the full range of six underlying climate subfactors available at the time of writing, rather than the combined index. These include: 
\begin{inparaenum}[1)]
\item percentage of days in the month in which the high temperature exceeds 90\textsuperscript{th} percentile (T90);
\item percentage of days in the months in which the low temperature is below the 10\textsuperscript{th} percentile (T10);
\item percentage of days in the month in which the wind power exceeds 90\textsuperscript{th} percentile in the reference period (WP90);
\item maximum number of consecutive dry days during a year with precipitation less than 1mm (CDD);
\item monthly maximum consecutive five-day precipitation amount in a month (Rx5Day); 
\item sea level in mm.
\end{inparaenum}
The above subfactors capture changes in the frequency and duration of extreme temperatures (highs and lows separately), heavy precipitation, drought, strong wind, and changes in sea level, and thus are representative of the key impacts of climate on people and the economy \citep{ACI25}. 

Turning to the macroeconomic factors, annual data over the same period as the cause of death and ACI subfactors were sourced from the World Bank \citep{WB25}, and include the following: 
\begin{inparaenum}
\item[1)] unemployment; 
\item[2)] real interest rate (RIR), and 
\item[3)] the consumer price index (CPI). 
\end{inparaenum}
Together, these three metrics reflect broad scale macroeconomic changes.

Unsurprisingly, preliminary checks revealed high collinearity among all the explanatory variables. For instance, T10 and T90 were highly negatively correlated, sea level and T90 were strongly positively correlated, and sea level and T10 were strongly negatively correlated; see Figure~\ref{fig:correlation-matrix} in Appendix~\ref{app:EDA} for a full correlation matrix. Broadly speaking, these correlations suggest extreme cold temperatures were linearly associated with windy days, and moderately associated with higher interest rates and CPI. Conversely, lower temperatures were not linearly associated with extreme heat, high sea levels, and rainfall, while extreme heat could be associated with higher sea levels and rainfall. 

With the strong multicollinearity in mind, we applied PCA to reduce the nine climate and macroeconomic explanatory variables to a smaller number of major axes of variation, prior to inputting them into our CODA regression models \citep{JC16, GGH+22}. In particular, we focused on the first three principal components, as these collectively explained more than 80\% of the total variance \citep[noting this is a commonly used but \emph{ad-hoc} guideline for choosing the number of principal components e.g.,][]{JC16}; see also Table~\ref{tab:importance_pc} in Appendix~\ref{app:EDA} for a full summary of the proportion of variance explained by the PCA. 

Table~\ref{tab:pc_loadings} presents the estimated loadings associated with the first three principal components, noting that the first three principal components explain 52\%, 19\%, and 12\%, respectively. Of these, the first (PC1) captures the strong positive and negative contributions of temperature (T10 and T90) variables, as well as the positive contribution of sea level. In particular, increasing values of PC1 correspond to higher T90 (hot days), higher sea levels, and lower T10 (cold extremes), thus reflecting warmer overall conditions with higher sea levels and fewer extreme cold events. Given the known differences in heat- and cold-related deaths \citep{GGH+15}, we anticipate that both T10 and T90 will be pivotal for forecasting cause-specific mortality. The second principal component was strongly correlated with CDD and Rx5Day (another variable for precipitation). Specifically, increasing PC2 corresponds to fewer consecutive dry days and more extreme rainfall events, thus representing wetter conditions with more intense precipitation, and suggesting that periods with high consecutive dry days during a year are often associated with periods of high precipitation over a consecutive five-day period. Finally, the third principal component increases with higher CPI and lower unemployment, and increases moderately with lower WP90 (wind power). As such, PC3 combines macroeconomic effects with wind-extreme effects, such that higher values  correspond to higher inflation, lower unemployment, and moderately lower high-wind effects. Put simply, PC3 reflects greater socioeconomic stress along with lower wind extremes.

\begin{table}[htb]
\centering
\begin{tabular}{@{}lcccl@{}}
\toprule[1.5pt]
\textbf{} & PC1 & PC2 & PC3 & \\
\textbf{} & (51.72\%) & (18.81\%) & (11.92\%) & \text{Commentary} \\
\cmidrule{2-5}
\emph{Climate subfactors} & & \\
\text{T10}            & \textbf{-0.4327} &  0.1245 &  0.1829 & Negative contributor to PC1 \\
\text{T90}            & \textbf{0.4397}  & -0.0109 &  0.1214 & Positive contributor to PC1 \\
\text{CDD}            &  0.1438 & \textbf{-0.6278} & -0.2143 & Negative contributor to PC2 \\
\text{Rx5Day}         &  0.3142 & \textbf{0.4781}  & -0.0428 & Positive contributor to PC2 \\
\text{sea level}       & \textbf{0.4381}  &  0.1486 &  0.0752 & Positive contributor to PC1 \\
\text{WP90}           & -0.3241 &  0.2570 & -0.3184 & -- \\\\
\emph{Macroeconomic factors} & & \\
\text{Unemployment}   & -0.1607 & -0.3113 & \textbf{-0.5282} & Negative contributor to PC3 \\
\text{RIR}            & -0.3721 &  0.2627 &  0.0252 & -- \\
\text{CPI}            & -0.1969 & -0.3280 & \textbf{0.7193}  & Positive contributor to PC3 \\
\bottomrule[1.5pt]
\end{tabular}
\caption{Estimated loading matrix for the three principal components, based on applying PCA to US explanatory variables. Bold values denote estimated loadings with absolute magnitude greater than 0.40, while the percentage of variance explained by the first three components is shown in parentheses at the top of the table.}\label{tab:pc_loadings}
\end{table}

\section{Statistical analysis}\label{sec:methodology}

We first establish some notation for cause of death responses and explanatory variables, before discussing the CODA framework (Section \ref{subsec:coda}) and two regression models fitted within this (Section \ref{subsec:regressionmodels}). 
 
Cause-specific mortality can be represented by actual death counts per combination of year, age band, and cause. Specifically, let $D_{t, u, c}$ denote the actual death count for year~$t = 1, 2, \dots, T$, age band $u = 1, 2, \dots, U$, and cause $c = 1, 2, \dots, C$, and define $D_t = ~\sum_{u=1}^U \sum_{c=1}^C D_{t, u, c}$ as the total deaths across all age bands and cause groups for year~$t$. Note that in the case of the US data, $T = 45$ years, $U = 9$ age bands, and $C = 10$ cause groups (including ``Other"). It follows that we can calculate $d_{t, u, c} = D_{t, u, c}/D_{t}$ such that, for a given year, the compositional vector 
 \begin{align} 
\bm{d}_t = (d_{t,1,1},d_{t,1,2},\ldots,d_{t,1,C},d_{t,2,1},d_{t,2,2}, \ldots, d_{t,2,C}, d_{t,u,1},d_{t,u,2}, \ldots,d_{t,U,C}) 
 \end{align}
represents the density of deaths by age band and cause, where we have ordered the elements such that cause runs faster than age. Clearly, the vector satisfies $\sum_{u=1}^U \sum_{c=1}^C d_{t, u, c} = 1$. 
Equivalently, due to the sum-to-one constraint, only $UC-1$ elements are needed to uniquely determine each vector $\bm{d}_t$. The sample space for compositional cause of death mortality data is thus a simplex, such that for all $t = 1, \dots T$ we have
 \begin{align} 
S^{UC-1} = \{ (d_{t, 1, 1}, \dots, d_{t, U, C}) | d_{t, u, c} \geq 0,\quad \sum_{u = 1}^{U} \sum_{c = 1}^{C} d_{t, u, c} = 1 \}.
 \end{align}

By stacking the $\bm{d}_t$'s as row vectors on top of each other, we can write the $T \times UC$ compositional matrix $\mathbf{D}$ of death densities as
\begin{equation}\label{eqn:D Matrix}
\mathbf{D} = \begin{pmatrix}
d_{1, 1, 1} & d_{1, 1, 2} & \dots & d_{1, 1, C} & d_{1, 2, 1} & d_{1, 2, 2} & \dots & d_{1, U, C} \\
d_{2, 1, 1} & d_{2, 1, 2} & \dots & d_{2, 1, C} & d_{2, 2, 1} & d_{2, 2, 2} & \dots & d_{2, U, C} \\
\vdots & \vdots & \ddots & \vdots & \vdots & \vdots & \ddots & \vdots \\
d_{T, 1, 1} & d_{T, 1, 2} & \dots & d_{T, 1, C} & d_{T, 2, 1} & d_{T, 2, 2} & \dots & d_{T, U, C} \\
\end{pmatrix}.
\end{equation}

At each time point, we also let $\bm{x}_t = (x_{t1}, x_{t2}, \ldots, x_{tp})$ denote the corresponding vector of $p$ covariates, e.g., the three principal components discussed in the preceding section, or basis functions formed from these principal components, as in the case of GAMs.

\subsection{Compositional data analysis} \label{subsec:coda}

We apply the CODA framework to account for the compositional constraint in the responses $\bm{d}_t$. At its core, this requires transforming compositional data from the simplex to real space, where it is unconstrained and circumvents issues with the coherence of the explanatory variables \citep{A86}. After fitting a regression model to the transformed responses, the results are then back-transformed to the compositional space for interpretation, inference, and prediction. 

By far, the two most common transformations used for CODA are the centred log-ratio (CLR) and the isometric log-ratio (ILR). The former 
is defined as dividing all values in the compositional vector by their geometric mean, before applying the natural logarithm. Mathematically, for row $t$ in~\eqref{eqn:D Matrix}, the CLR for each element is given by
\begin{align}
w(d_{t, u, c}) &= \ln\left(\frac{d_{t, u, c}}{(\prod_{u=1}^U \prod_{c=1}^C d_{t, u, c})^{1/UC}}\right) = \ln(d_{t,u,c}) - \frac{1}{UC} \sum_{u=1}^U\sum_{c=1}^C \ln(d_{t,u,c}).\label{eqn:CLR}
\end{align}
The CLR transformation is symmetric relative to the compositional parts, with the resulting transformed vector $\bm{w}(\bm{d}_t) = (w(d_{t,1,1}),w(d_{t,1,2}),\ldots,w(d_{t,U,C}))$ defined such that the distance between any two elements of this vector remains the same when measured in the simplex and the real space.
On the other hand, while each element of $\bm{w}(\bm{d}_t)$ is no longer constrained to be non-negative, the entire vector remains constrained since the elements must sum to zero by the construction of~\eqref{eqn:CLR}.
To further remove this constraint, the ILR transformation left multiplies $\bm{w}(\bm{d}_t)$ by a orthonormal $(UC-1) \times UC$ Helmert sub-matrix formed by deleting the first row of the Helmert orthogonal matrix \citep[see][for technical details]{G21, TS22}. Denoting this Helmert sub-matrix as $\bm{H}$, the ILR-transformed vector is given by
\begin{equation}\label{eqn:ILR}
    \bm{z}(\bm{d}_t) = \bm{H} \bm{w}(\bm{d}_t),
\end{equation}
which is no longer subject to any constraint, i.e., $\bm{z}(\bm{d}_t) \in \mathcal{R}^{UC-1}$. The ILR is often promoted as the more theoretically correct method, especially for contrasting groups of elements, in the CODA framework \citep{G21}. 

In many death count by cause responses, such as the US population data at the centre of the analysis in this article, zero counts occur for certain combinations of years, age bands, and causes. This poses issues when both transformations \eqref{eqn:CLR} or \eqref{eqn:ILR} above are used, due to having to take the logarithm at zero. Typical \emph{ad-hoc} workarounds to this include collapsing combinations, or adding a small arbitrary value to move values away from the zero boundary. In this article, however, we adopt an alternative, statistically more principled approach in the form of the $\alpha$-transformation, an extension of the ILR that can be viewed as a type of Box-Cox transformation applied to the ratios of components \citep{TPW11}. Briefly, for row $t = 1,\ldots,T$ in~\eqref{eqn:D Matrix}, the $\alpha$-transformation is defined as:
\begin{equation}\label{eqn:alpha}
    \bm{z}^\alpha(\bm{d}_t) = \bm{H} \bm{w}^\alpha(\bm{d}_t),
\end{equation}
where 
$\bm{w}^\alpha (\bm{d}_t)$ denotes the vector where the Box-Cox transformation \citep{BC64} is applied element-wise to $\bm{d}_t$. That is, $w^\alpha(d_{t,u,c}) = \ln(d_{t,u,c})$ if $\alpha = 0$; otherwise $w^\alpha(d_{t,u,c}) = (UC)(d_{t,u,c}^\alpha -1)/\alpha$ for $\alpha \in (0,1]$. 
It is not hard to see that when $\alpha = 0$, the $\alpha$-transformation reduces to the ILR transformation. More importantly, when $\alpha$ is restricted to be greater than zero, the transformed values are well defined even when one or more $d_{t,u,c} = 0$; see \citet{DSH+25} for further details of the $\alpha$-transformation, particularly in the context of mortality forecasting. Statistically, the corresponding sample space of the $\alpha$-transformation is known as the $\alpha$ space, which is given formally by
$\mathbb{A}^{UC-1}_\alpha = \{ \bm{z}^\alpha(\bm{d}_t) | -\alpha^{-1} \leq w^\alpha(d_{t,u,c}) \leq (UC-1)^{-1}\alpha; \sum_{u=1}^{U}\sum_{c=1}^{C} w^\alpha(d_{t,u,c}) = 0 \}$.
As $\alpha \rightarrow 0$, then $\mathbb{A}^{UC-1}_\alpha$ tends to the real space $\mathcal{R}^{UC-1}$; this is consistent with the ILR, except now zero values of death densities can be handled, provided $\alpha \ne 0$.

While the turning parameter $\alpha$ is often determined using a data-driven approach via maximum likelihood estimation \citep{TPW11}, minimising some form of the sum of squared errors \citep[e.g.,][]{TA25}, or by cross-validation \citep{DSH+25}, in this article we fix $\alpha = 0.5$ for consistency and ease of interpretation. Other values for $\alpha$ were tested, but the overall impacts on the scenario analysis below relatively small.

\subsection{CODA regression models}\label{subsec:regressionmodels}

Having applied the $\alpha$-transformation in~\eqref{eqn:alpha} to the compositional matrix of death densities $\bm{D}$,  
we apply either a multivariate linear regression model or the more flexible multivariate GAM on the $\alpha$-transformed responses. Such CODA regression approaches, or some variation thereof, have been applied to a number of other fields, such as biology, health economics, and geology \citep[e.g.,][]{HS23}, although their usage in the actuarial sciences is relatively new. For the remainder of this article, we will refer to the above two methods as CODA-LM and CODA-GAM, respectively. 
In the former, we have
\begin{equation*}
\bm{z}^\alpha(\bm{d}_t) = \bm{B}^\top \bm{x}_t + \bm{\varepsilon}_t, 
\quad \bm{\varepsilon}_t \sim \mathcal{N}(0, \bm{\Sigma}); \quad t = 1,\ldots,T,
\end{equation*}
which can be written in matrix form simply as $\bm{Z} = \bm{X} \bm{B} + \bm{E}$, where $\bm{Z}$ is the $T \times (UC-1)$ matrix formed by stacking the $\bm{z}^\alpha(\bm{d}_t)$ on top of each other as row vectors, and similarly for the $T \times p$ matrix $\bm{X}$ formed from the $\bm{x}_t$'s, $\bm{B}$ denotes a corresponding $p \times (UC-1)$ coefficient matrix, and $\bm{E}$ denotes the $T \times (UC-1)$ error matrix formed by stacking the $\bm{\varepsilon}_t$'s. We assume $\bm{\Sigma}$ is a $(UC-1) \times (UC-1)$ diagonal matrix of error variances. 

With CODA-LM, we set up $\bm{x}_t$ to comprise an intercept as its first element, along with the three principal components derived in Section \ref{subsec:explanatorydata} which we include as linear terms. CODA-GAM modifies this such that $\bm{x}_t$ involves smooth terms formed using basis functions from the three principal components, which are then coupled with one or more smoothing penalty matrices used in the estimation process to penalise excessive wiggliness \citep{mgcv}. Doing so allows for flexible non-linear, but additive, relationships between the transformed cause of death densities and the climate/macroeconomic factors. We refer to \citet{mgcv} for details for the latter are implemented, noting for our analysis we used the default penalised thin plate splines available in \Rlogo\ package \texttt{mgcv} \citep{mgcv}. We also tested other methods for constructing the splines and penalty functions, but found that the overall conclusions below varied little with these choices.

Note that in constructing $\bm{\Sigma}$ to be a diagonal matrix, both CODA-LM and CODA-GAM assume independence across all components in the composition. That is, there is no residual correlation between the $(UC-1)$ elements of $\bm{z}^\alpha(\bm{d}_t)$ in the transformed $\alpha$-space. This is a potentially strong assumption for CODA regression, and future research could examine relaxing it, such that $\bm{\Sigma}$ can capture potential residual covariation above and beyond that due to the principal components \citep[noting a key challenge here will be handling the number of parameters in $\bm{\Sigma}$ relative to $T$; see][for recent examples of potential workarounds]{hui2024homogeneity,gao2025penalized}.  

After fitting either the CODA-LM or CODA-GAM above, we construct fitted values and predictions of the compositions of deaths by cause and age band, denoted here by $\bm{z}^\alpha(\widehat{\bm{d}_t})$, by performing the inverse $\alpha$-transformation back to the simplex, i.e., $\widehat{\bm{d}}_t = {\alpha}^{-1}(\bm{z}^\alpha(\widehat{\bm{d}}_t))$. Uncertainty intervals around these fitted values and predictions can be obtained in an analogous manner based on intervals obtained from the multivariate linear model or GAM.

\section{Cause of death data analysis}\label{sec:application}
We applied CODA-LM and CODA-GAM to the US population data, for males and females separately, before using the latter, in particular, to perform scenario and sensitivity analysis and provide insight into the associations between climate-related factors and specific causes of death. Section~\ref{sec:model_performance} focuses on the goodness of fit for the CODA-LM and CODA-GAM models, 
Section~\ref{sec:model_results} delves into the details of the modelled results, and Section~\ref{sec:scenarios} explores the scenario application of the CODA-GAM model.

\subsection{Model performance}\label{sec:model_performance}
To evaluate the model fit and compare performance between CODA-LM and CODA-GAM, we computed both the $\alpha$-space $R^2$ and Aitchison's $R^2$ averaged across the $U \times C$ fitted values.
Briefly, the $\alpha$-space $R^2$ measures the goodness of fit relative to the reference model, i.e., the alpha-mean composition, in the $\alpha$-transformed real space, and quantifies how well the model explains variation in the transformed compositions. Note that its value is dependent on the chosen $\alpha$ and thus reflects the geometry induced by that transformation. 
By comparison, Aitchison's $R^2$ measures explained variability with respect to log-ratio distances in the constrained simplex, 
meaning relative changes among components are emphasised, particularly when components vary by several orders of magnitude. 
As a result, Aitchison's $R^2$ is said to provide a more interpretable, geometry-consistent summary of explanatory power that closely aligns with the fundamental principles of compositional data analysis \citep{A86}.


Perhaps unsurprisingly, CODA-GAM achieved higher values of $R^2$ compared with CODA-LM, due to (expected) non-linear relationships between climate factors and cause of death mortality and the capacity of the former to better account for this. In particular, for US females, the $\alpha$-space $R^2$ was 0.9558 (0.8739 for the equivalent using CODA-LM) and the Aitchison $R^2$ is 0.9971 (0.9944 for the equivalent using CODA-LM), while for US males the $\alpha$-space $R^2$ was 0.9486 (0.8419 for the equivalent using CODA-LM) and the Aitchison $R^2$ was 0.9823 (0.9815 for the equivalent using CODA-LM). Note that all the actual values of $R^2$ are relatively close to one, meaning the models performed well in capturing how the causes of death increase or decrease in an absolute sense. This also suggests that the effects of the principal components could be reasonably included in an additive manner. 
In Appendix~\ref{sec:Appendix_US_GAM_Fit}, we provide further details on the computed $R^2$ values broken down by each of the components in the composition, along with 
plots of the fitted CODA-GAM against true values, for the top ten cause and age bands.

\subsection{Model results}\label{sec:model_results}

In light of the above results, for the remainder of the application, we focus on the fitted CODA-GAM models given the clear evidence of non-linear effects of climate factors on mortality by cause and age; see Table~\ref{fig:LMCODA_US_Results_Commentary} in Appendix~\ref{sec:coda-lm} for a brief summary of the results from CODA-LM for completeness.

Note that, as with all CODA based methods, the inverse transformation does not directly recover the original cause-age labels associated with each fitted effect. Therefore, after applying the inverse $\alpha$-transformation, each predicted series is then reassigned to a cause-age combination by comparing it with the original transformed series. Specifically, for each inverse-transformed prediction, we compute the correlation between its fitted trajectory over time and the trajectories of all original cause–age combinations, and assign the label corresponding to the highest correlation \citep[see][for further details of this approach]{TPW11}.
In some cases, multiple cause–age combinations exhibit very similar temporal behaviour (e.g., neighbouring age bands within the same disease category). When this occurs, their transformed series are highly correlated, and so different components of the fitted CODA-GAM model may be matched to the same cause–age combination under the correlation-based reassignment procedure.
For instance, below, this explains why the US Female I033 Heart Ischaemic series for age 55 is not uniquely mapped among the top ten combinations: the temporal pattern is extremely similar to the adjacent Heart Ischaemic age groups, resulting in correlations that are nearly indistinguishable across those neighbouring bands. Consequently, another neighbouring age band achieves a slightly higher correlation and is selected during reassignment.

\begin{figure}[!htb]
\centering
\begin{subfigure}{0.45\textwidth}
\includegraphics[width=\linewidth]{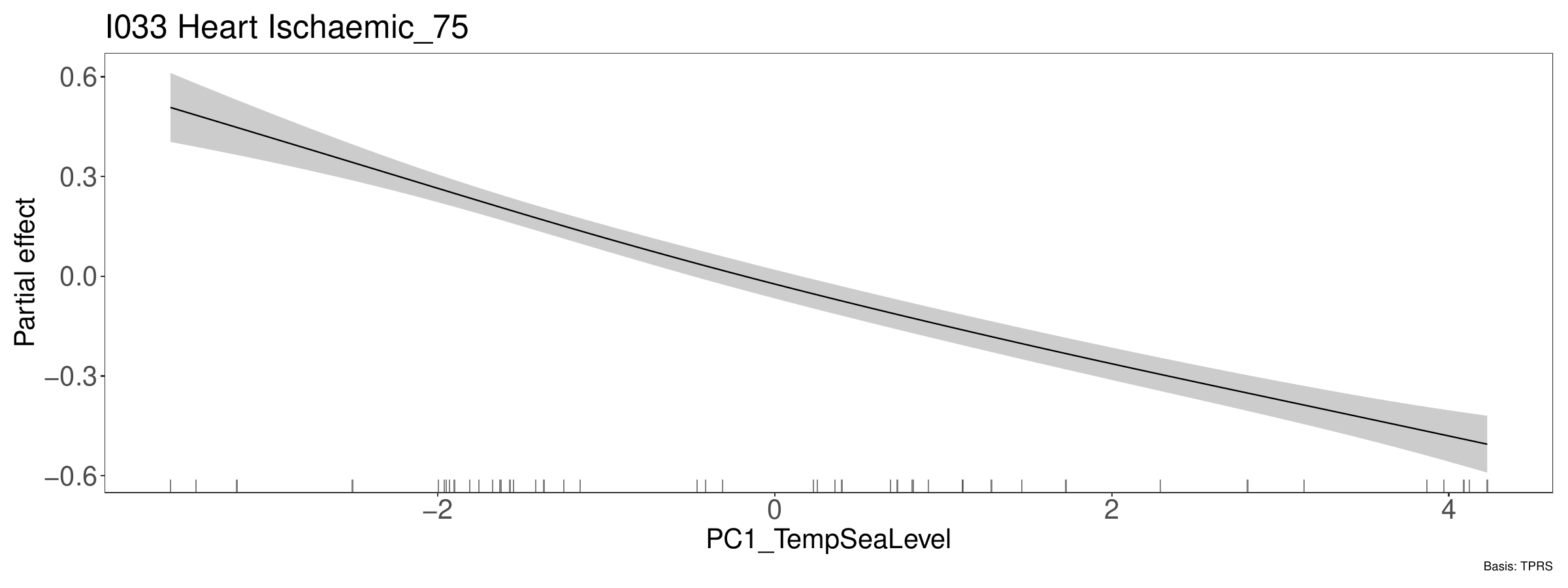}
\end{subfigure}
\begin{subfigure}{0.45\textwidth}
\includegraphics[width=\linewidth]{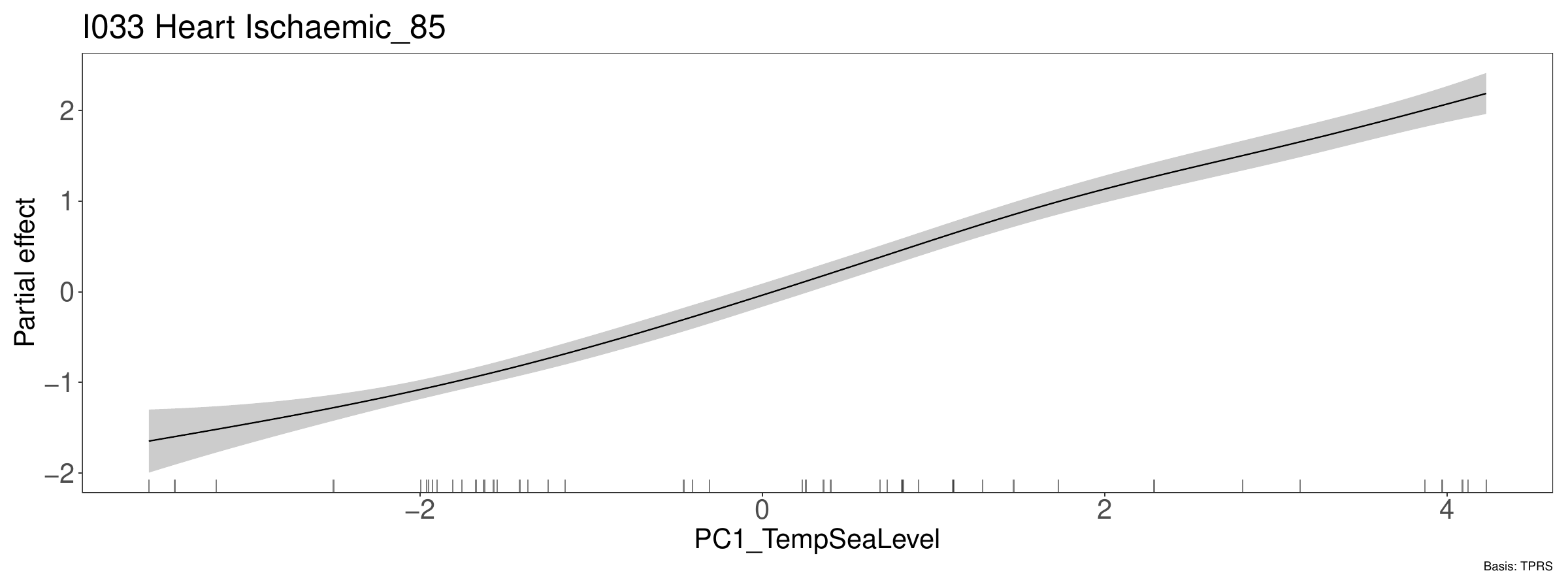}
\end{subfigure}

\begin{subfigure}{0.45\textwidth}
\includegraphics[width=\linewidth]{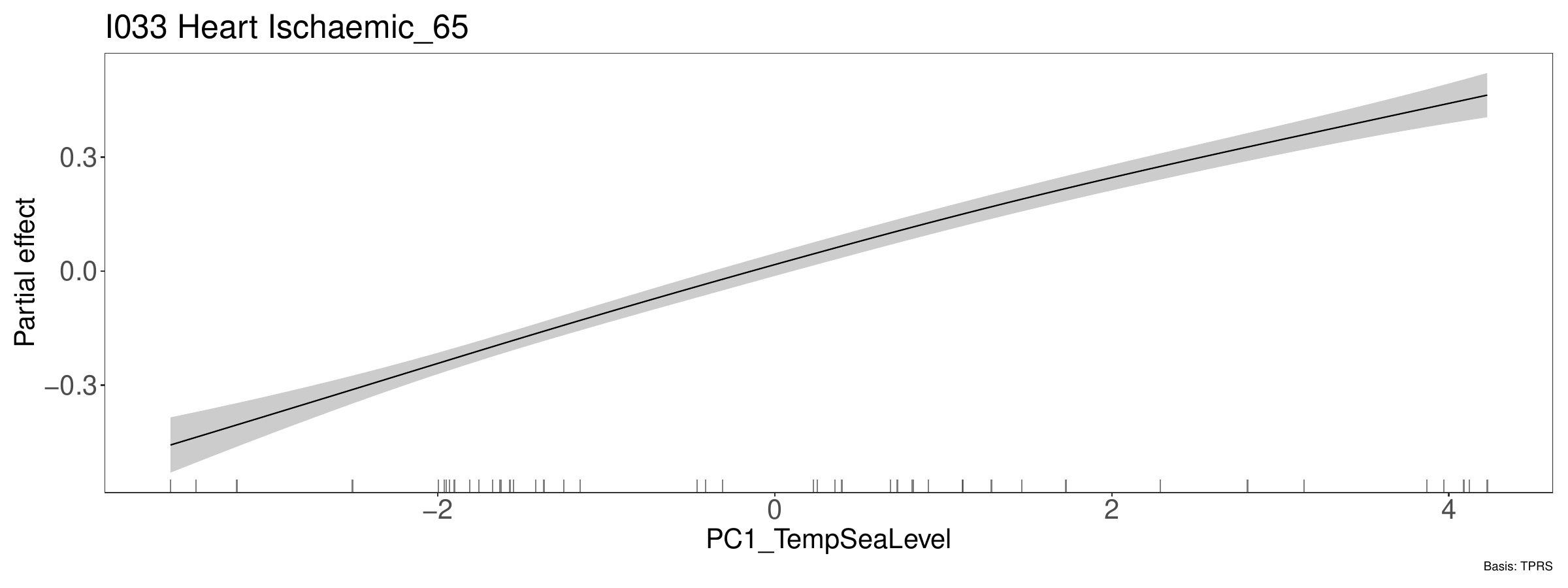}
\end{subfigure}
\begin{subfigure}{0.45\textwidth}
\includegraphics[width=\linewidth]{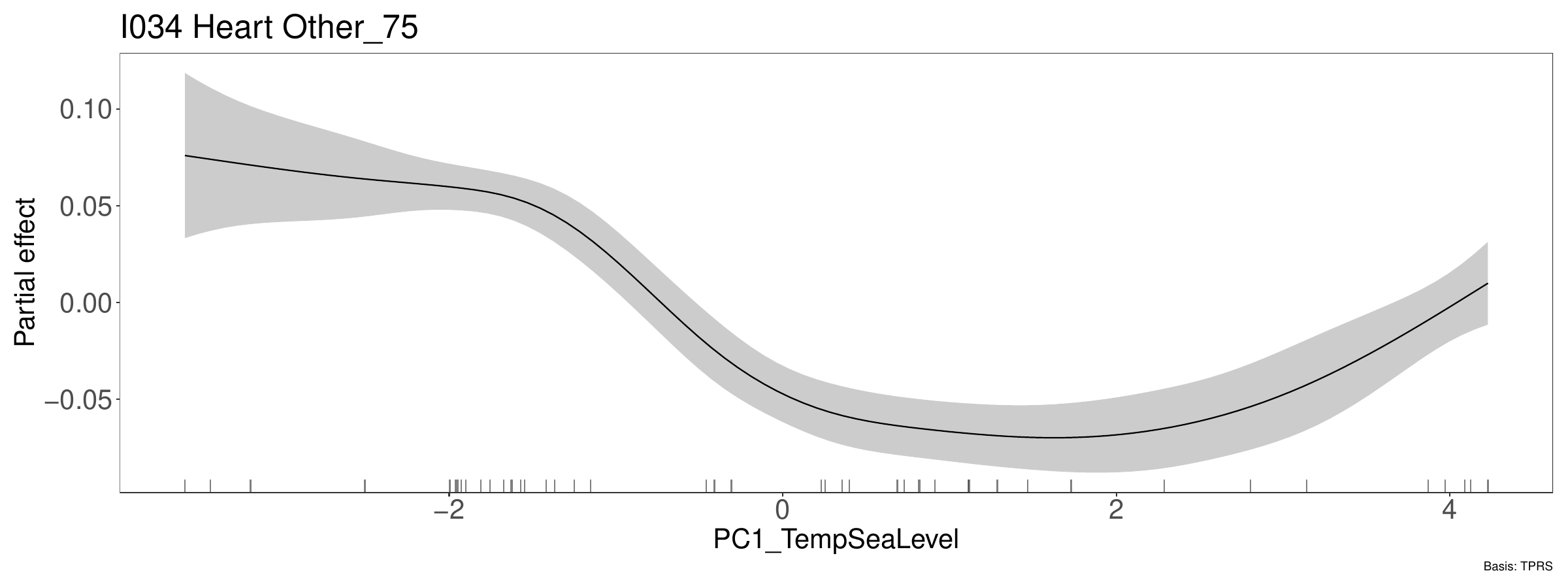}
\end{subfigure}

\begin{subfigure}{0.45\textwidth}
\includegraphics[width=\linewidth]{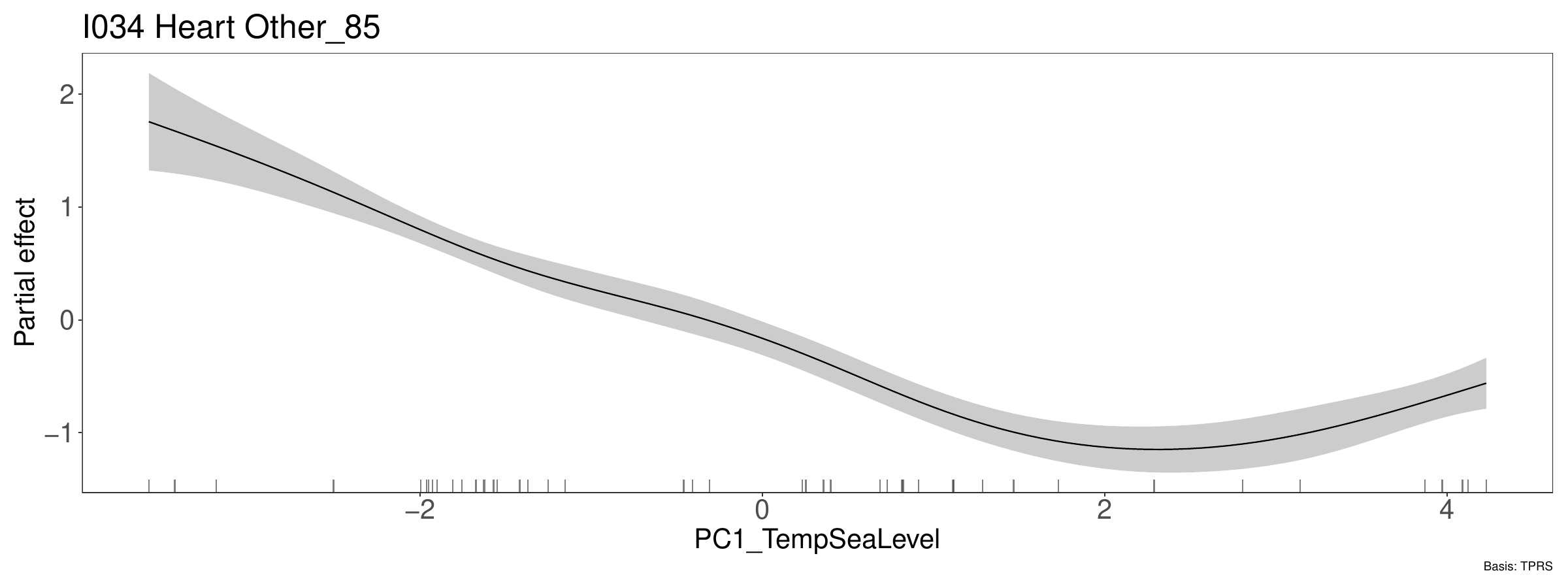}
\end{subfigure}
\begin{subfigure}{0.45\textwidth}
\includegraphics[width=\linewidth]{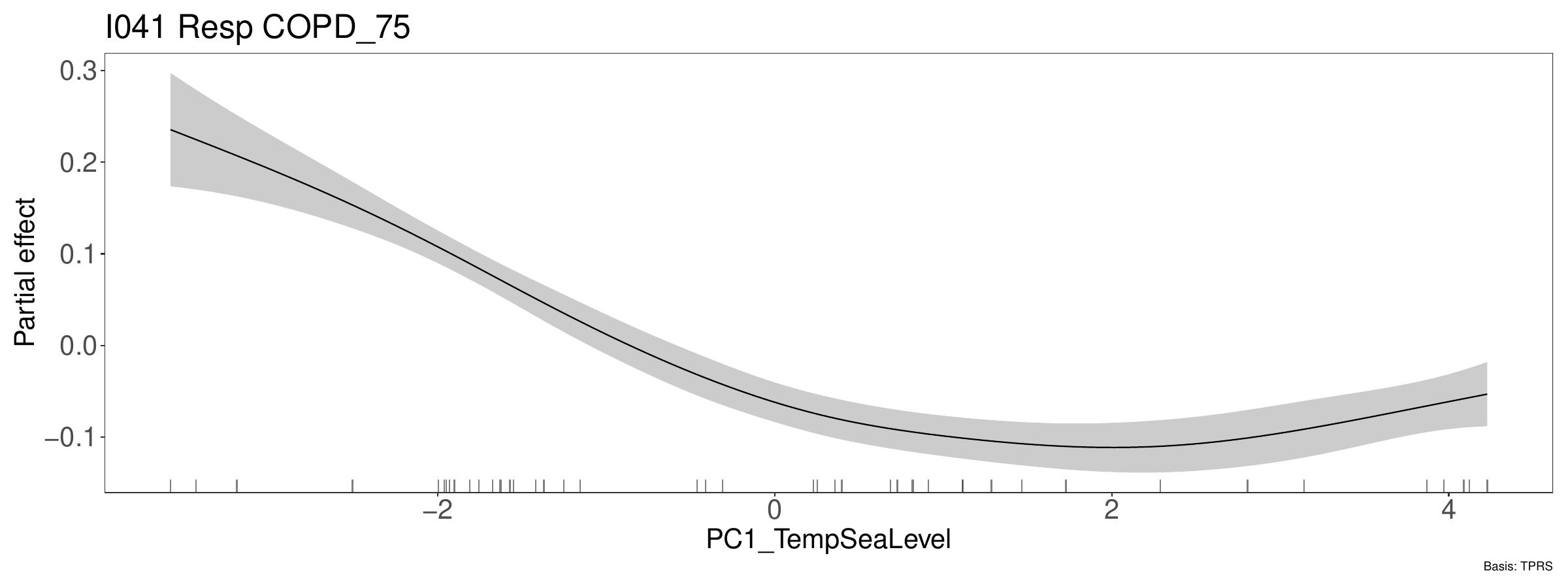}
\end{subfigure}

\begin{subfigure}{0.45\textwidth}
\includegraphics[width=\linewidth]{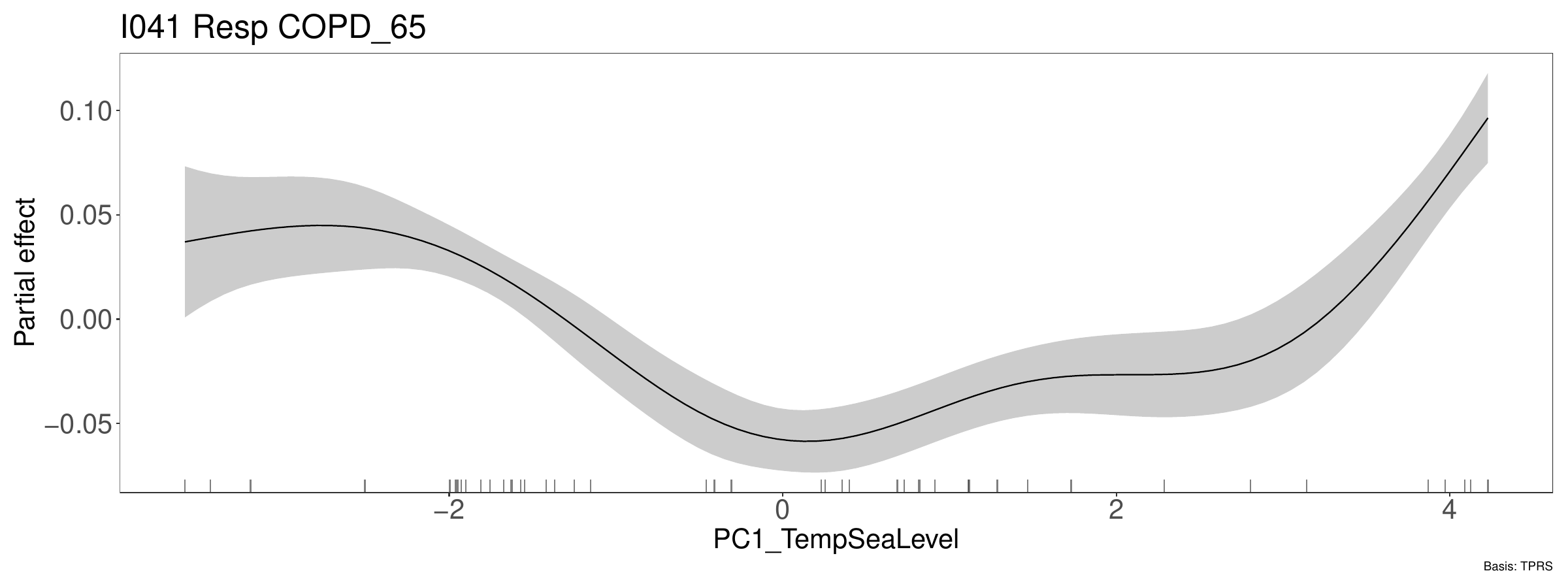}
\end{subfigure}
\begin{subfigure}{0.45\textwidth}
\includegraphics[width=\linewidth]{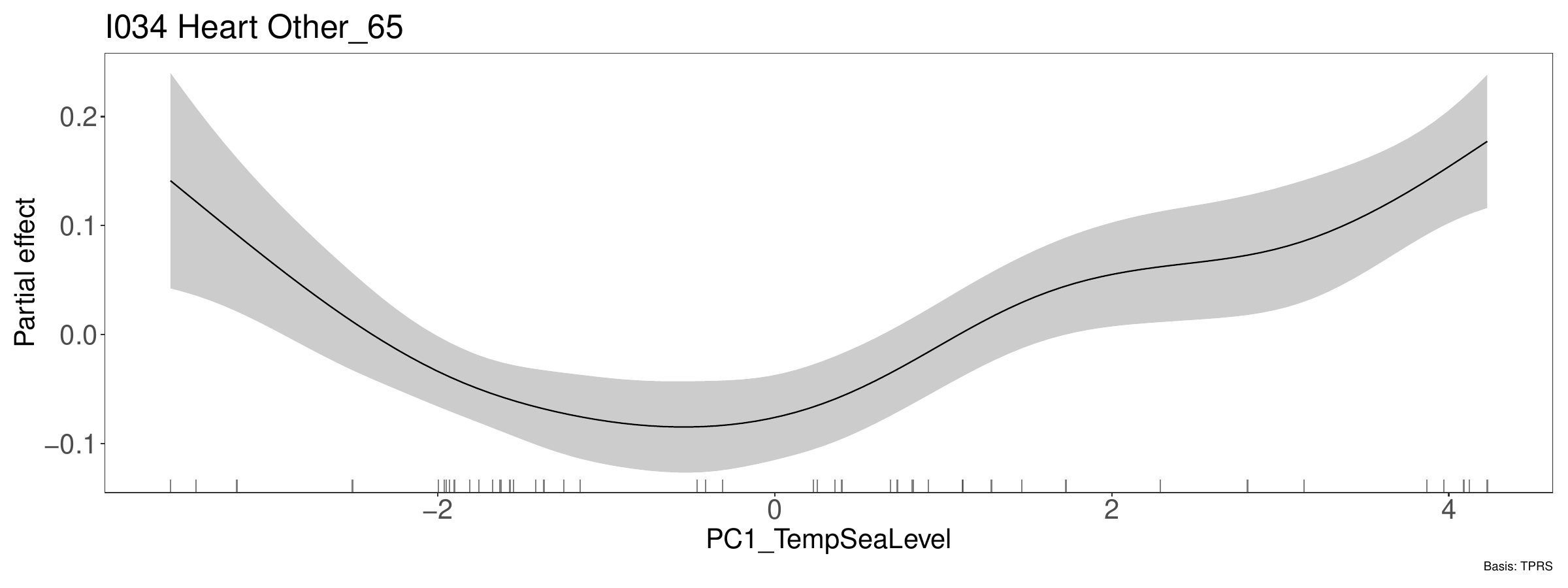}
\end{subfigure}

\begin{subfigure}{0.45\textwidth}
\includegraphics[width=\linewidth]{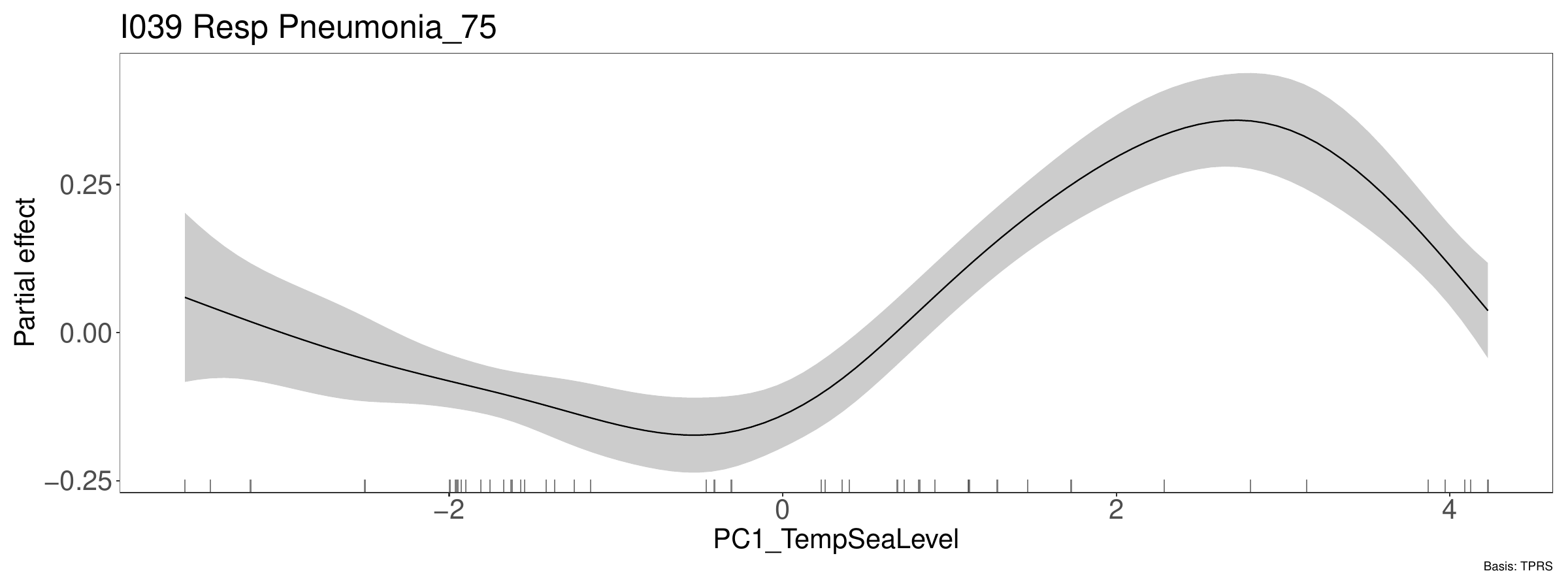}
\end{subfigure}
\caption{Partial effects plots of PC1 based on fitting a CODA-GAM to US females, and depicting the top cause and age-bands. Solid lines denote estimated partial effects (smooth term), while the shaded areas indicates the corresponding 95\% pointwise confidence band. Note that for US females, there are only 9 plots in the constrained compositional space as the transformed series are highly correlated.}
\label{fig:US_Partial_PC1_F}
\end{figure}

\begin{figure}[!htb]
\centering
\begin{subfigure}{0.45\textwidth}
\includegraphics[width=\linewidth]{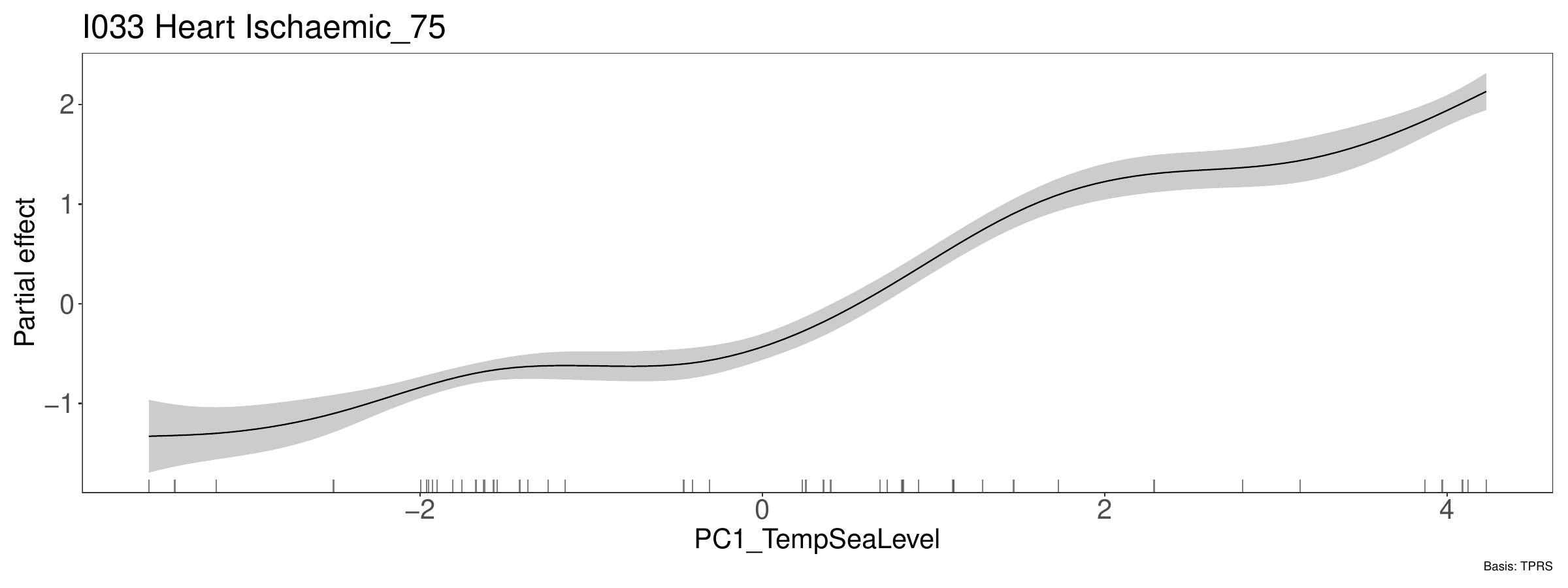}
\end{subfigure}
\begin{subfigure}{0.45\textwidth}
\includegraphics[width=\linewidth]{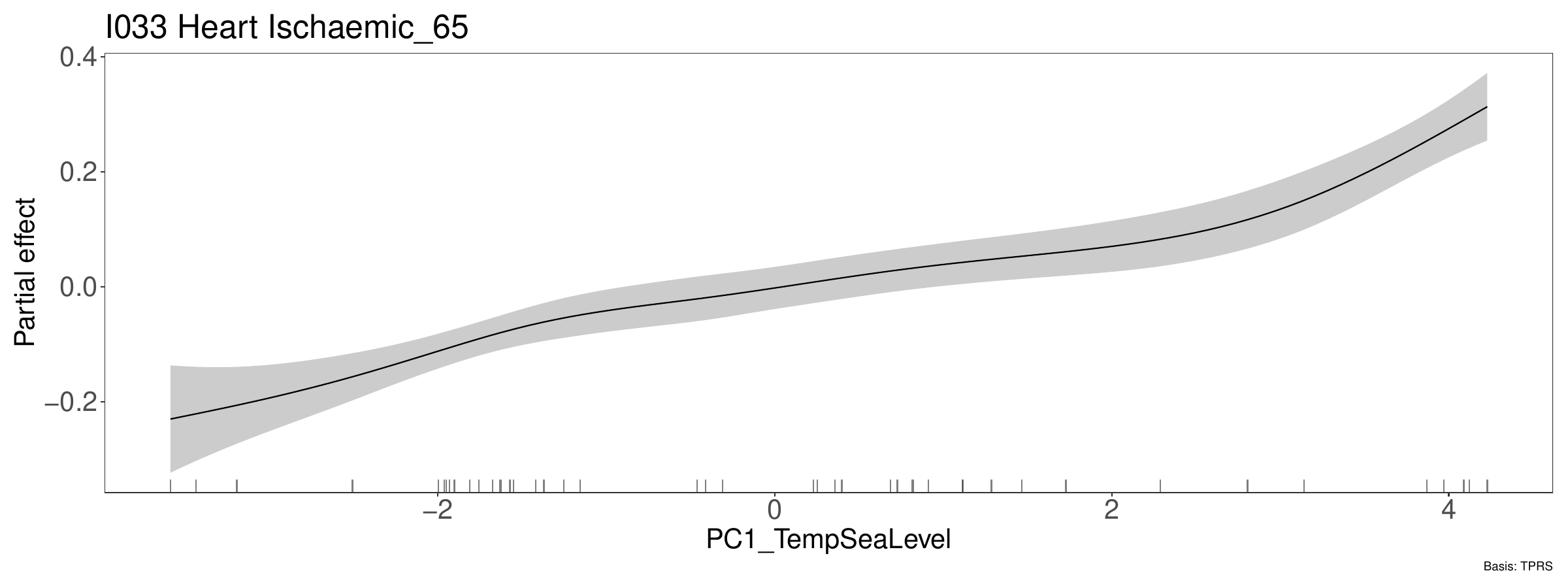}
\end{subfigure}

\begin{subfigure}{0.45\textwidth}
\includegraphics[width=\linewidth]{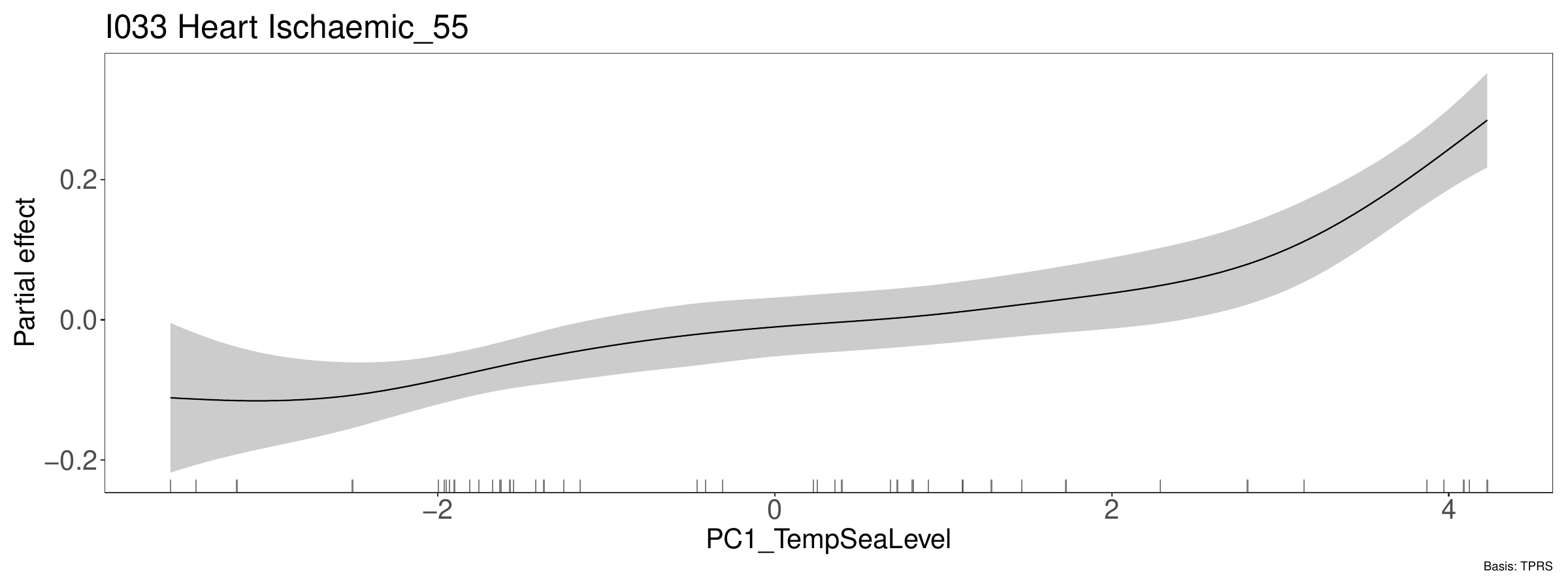}
\end{subfigure}
\begin{subfigure}{0.45\textwidth}
\includegraphics[width=\linewidth]{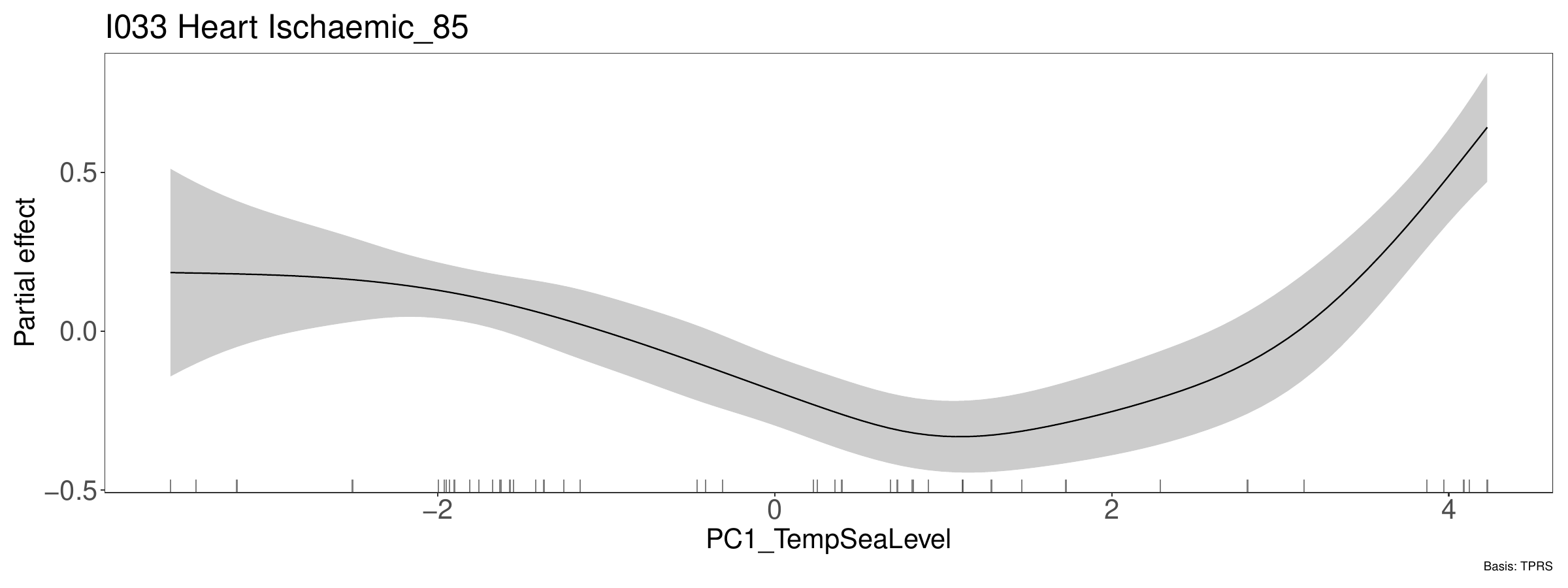}
\end{subfigure}

\begin{subfigure}{0.45\textwidth}
\includegraphics[width=\linewidth]{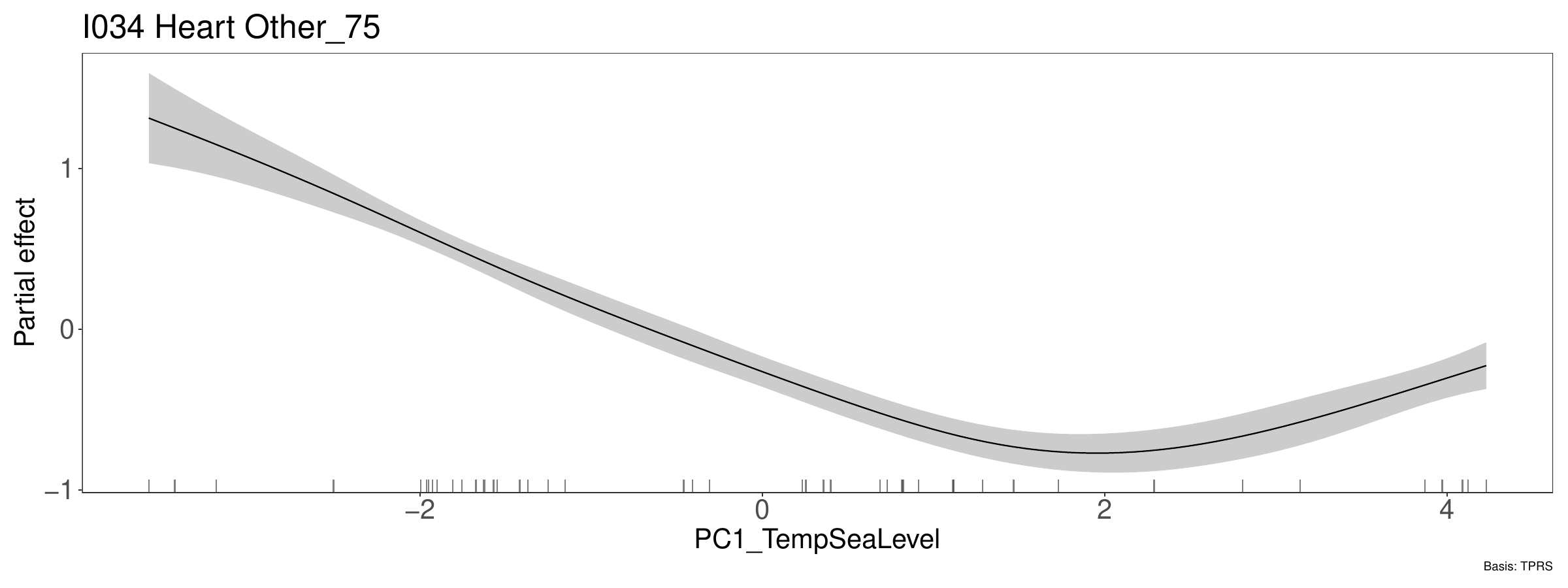}
\end{subfigure}
\begin{subfigure}{0.45\textwidth}
\includegraphics[width=\linewidth]{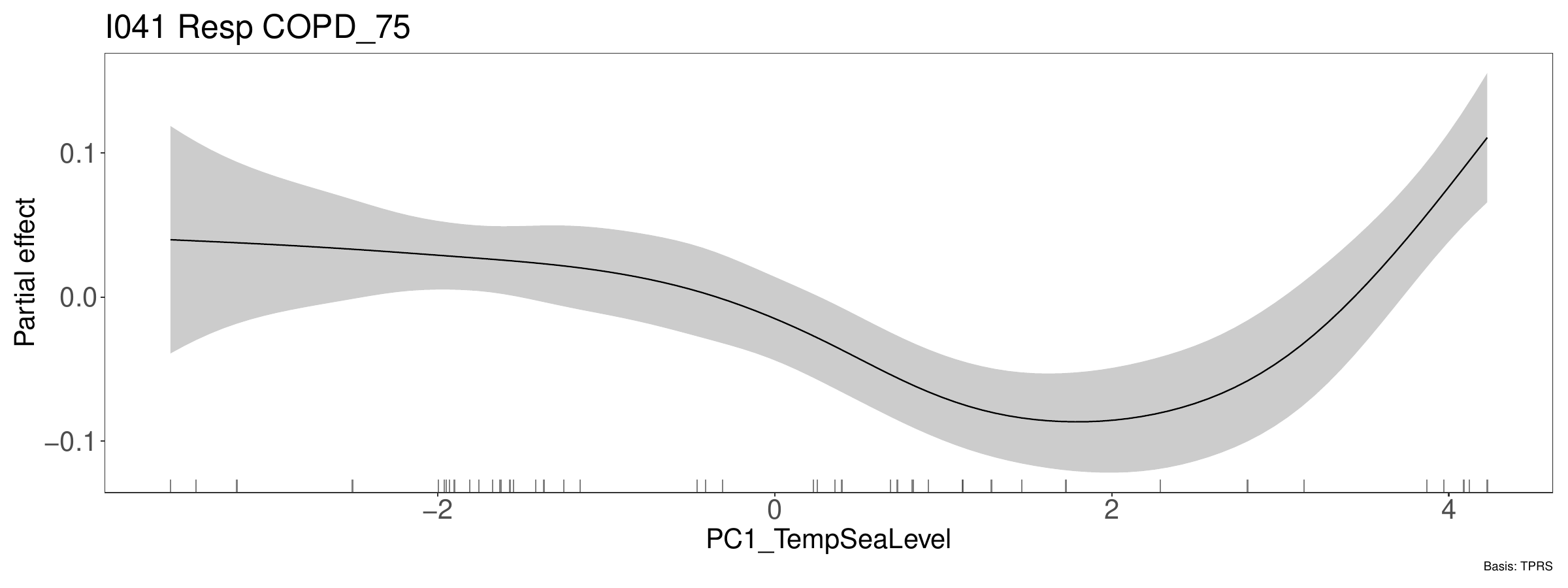}
\end{subfigure}

\begin{subfigure}{0.45\textwidth}
\includegraphics[width=\linewidth]{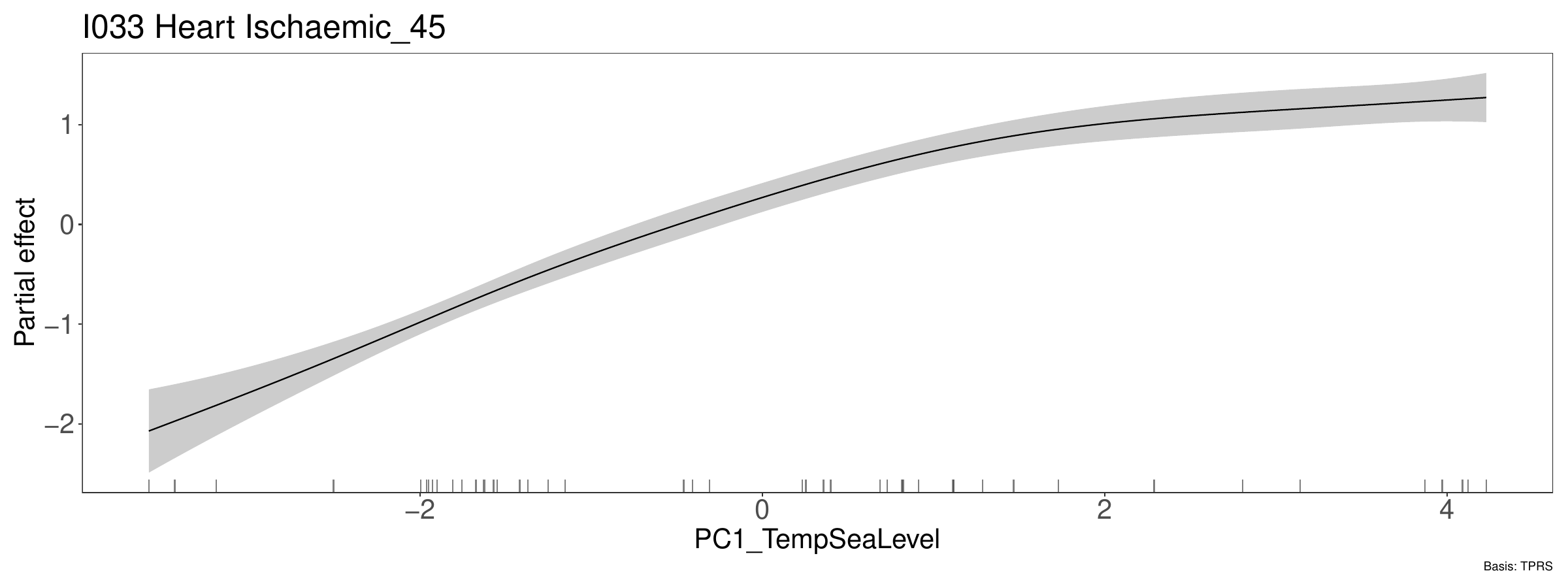}
\end{subfigure}
\begin{subfigure}{0.45\textwidth}
\includegraphics[width=\linewidth]{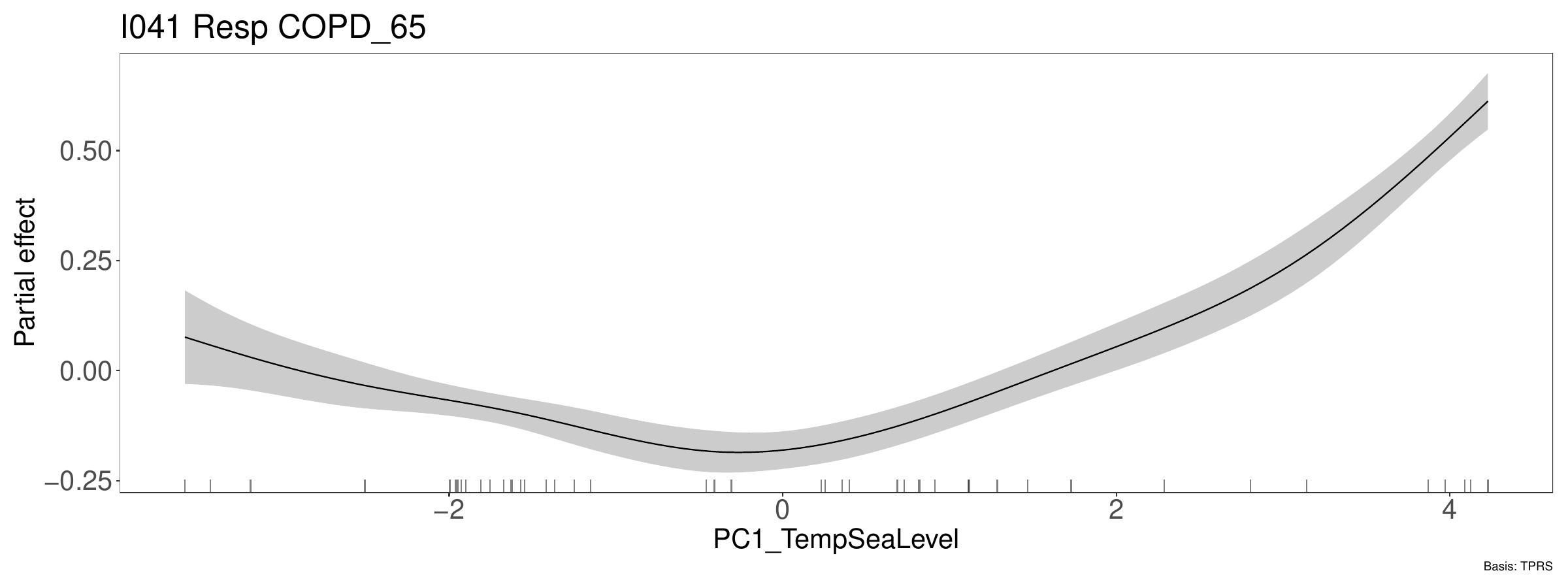}
\end{subfigure}

\begin{subfigure}{0.45\textwidth}
\includegraphics[width=\linewidth]{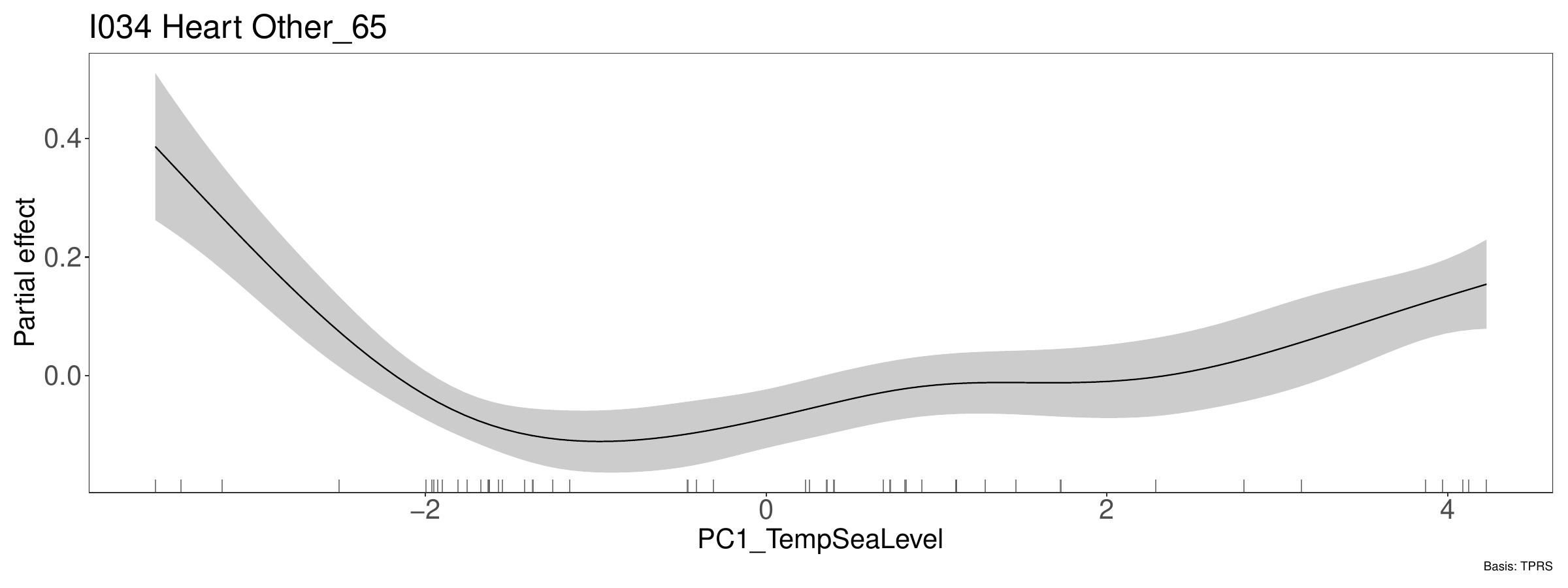}
\end{subfigure}
\begin{subfigure}{0.45\textwidth}
\includegraphics[width=\linewidth]{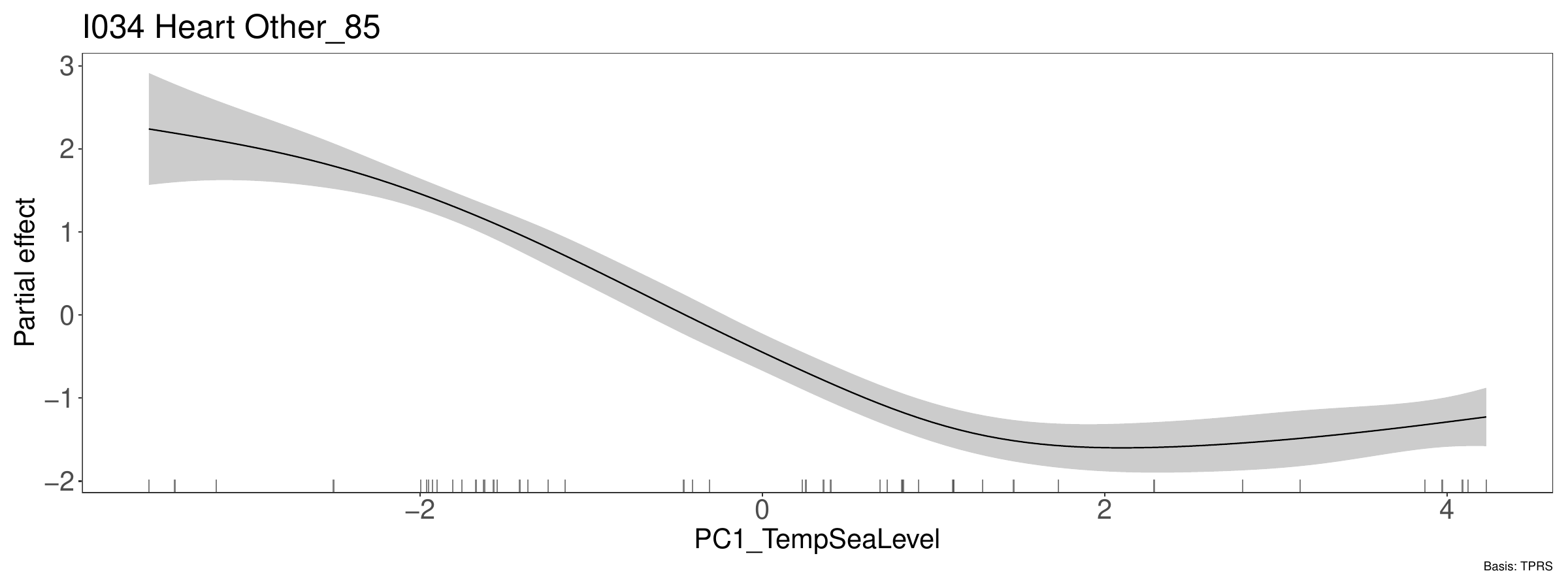}
\end{subfigure}
\caption{Partial effects plots of PC1 based on fitting a CODA-GAM to US males, and depicting the top ten cause and age-bands. Solid lines denote estimated partial effects (smooth term), while the shaded areas indicates the corresponding 95\% pointwise confidence band.}
\label{fig:US_Partial_PC1_M}
\end{figure}

Figures~\ref{fig:US_Partial_PC1_F} and \ref{fig:US_Partial_PC1_M} present the GAM partial plots with PC1 for females and males, respectively, where the solid line represents the estimated partial effect (smooth term) of the principal component on the response scale, and the grey shaded area is the corresponding 95\% pointwise confidence band around the smooth. 
Whilst the top three PCs were selected for modelling and used in CODA-GAM, we focus below on PC1 for interpretation, given that it includes the most relevant climate factors by capturing the combined temperature extremes and sea level effect and explaining more than 50\% of the variance. 

There is clear evidence of non-linearity for both the female and male US populations, where in the former, the non-linear association with PC1 is most prevalent in Heart Ischaemic at the oldest age band, as well as in Heart Other and Pneumonia. For US males, this is similar, though there is more non-linearity for respiratory COPD causes as well. 
The causes of ischaemic heart disease have fairly linear relationships for US females (Figure~\ref{fig:US_Partial_PC1_F}), with the proportion of deaths increasing linearly with increases in PC1 for the age bands 65-74 and 85-94, whereas the proportion of deaths decreases linearly with increases in PC1 for the 75-84 age band. This could indicate an acceleration of the deaths effect with increases in PC1, or other underlying risk drivers. For other cause and age combinations, the effect of heart other for ages 65-84, respiratory COPD 65-74 are similar, with all three groups showing a slight decrease in the proportion of deaths around the mean value of PC1, and increases at the extremes. This effect is somewhat present for the 85-94 age band for heart other, but with a greater increase at the highest levels of PC1. 
Turning to US males, Figure~\ref{fig:US_Partial_PC1_M} shows a similar increasing linear effect for Ischaemic heart disease for age bands 55-84. For ages 85-94 however, the Ischaemic heart disease cause appears to exhibit a slight U-shape, where the proportion of deaths decreases for values of PC1 until just above the historic mean, and then increases thereafter. This U-shape is also apparent in the respiratory COPD cause for ages 65-84, and to some extent for Other heart diseases at ages 65-74. We conjecture that the U-shape at moderate levels of PC1 may be driven by a reduction of deaths due to fewer cold extremes, which is consistent with findings per \citet{LKS+20}. The Other heart diseases cause for ages 85-94 suggests decreasing proportions of deaths for all values of PC1. This may be attributable to offsetting deaths at other age bands or for other causes, or an acceleration of deaths.


For completeness, we also include plots of the fitted CODA-GAM overlaid on the true values over the time series period for the top ten causes and age bands in Figure~\ref{fig:US_GAM_Fit} of Appendix~\ref{sec:Appendix_US_GAM_Fit}.

\subsection{Climate scenario and sensitivity analysis}\label{sec:scenarios}

We now apply CODA-GAM to understand the implications of climate change through scenario analysis. Specifically, we assess the effects on mortality by cause based on varying the PC1 over the range $-8.0$ to $8.0$. This particular range of values for the first principal component reflects an extrapolation equivalent to approximately double the historically observed range over the period of analysis. By testing further extremes, such counterfactual scenarios help to shed light on the compositional response for different causes of death and age bands beyond the historical data, in a more (or less) extreme climate. 

As reviewed in Section~\ref{subsec:litreview}, similar approaches to considering scenario analysis have been applied elsewhere. Moreover, the focus on varying PC1 again reflects the fact that it alone explains more than 50\% of the total variation across all explanatory factors, and it is most reflective of climate-related factors. In particular, recall from Table~\ref{tab:pc_loadings} that PC1 essentially reflects climate warming and intensification. 
Indeed, risk factors such as temperature extremes and sea levels are two of the factors considered in other climate scenarios, e.g., as part of the CBES scenarios \citep{BOE21}. 
By varying PC1 as per the above, we present a scenario indicative of more cold extremes and fewer hot extremes, coupled with a lower sea level (when PC1 is decreased to -8). Conversely, an increase in PC1 to 8 means more hot extremes, fewer cold extremes, and a higher sea level. The range of results is transformed back to the compositional space to infer the impact on the proportion of deaths.

\begin{figure}[!htb]
\centering
\includegraphics[width=15.5cm]
{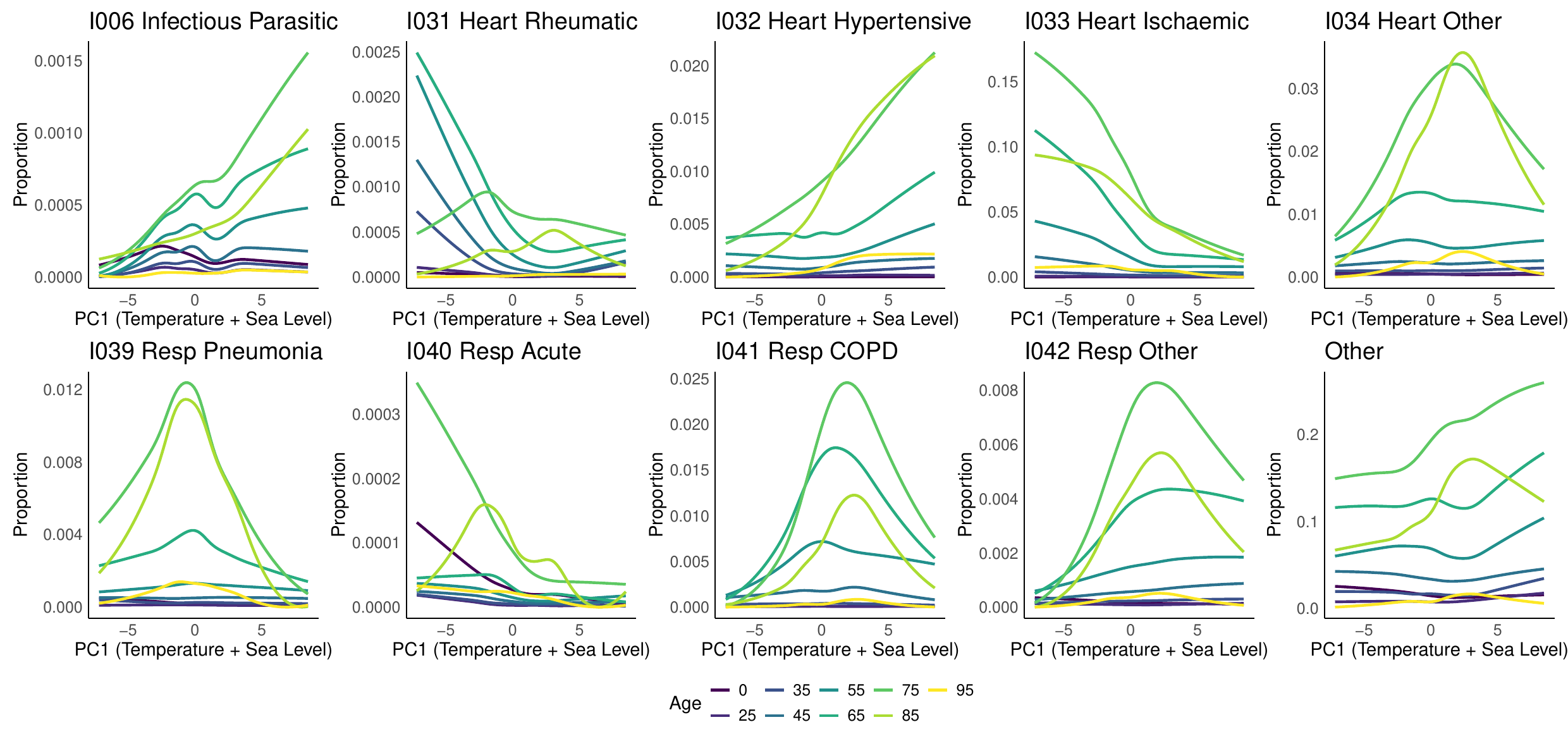}
\caption{Scenario analysis for US females, using the fitted CODA-GAM and based on varying PC1 from -8.0 to 8.0.}\label{fig:US_F_PC1_Sensitivity}
\end{figure}
\begin{figure}[!htb]
\centering
\includegraphics[width=15.5cm]
{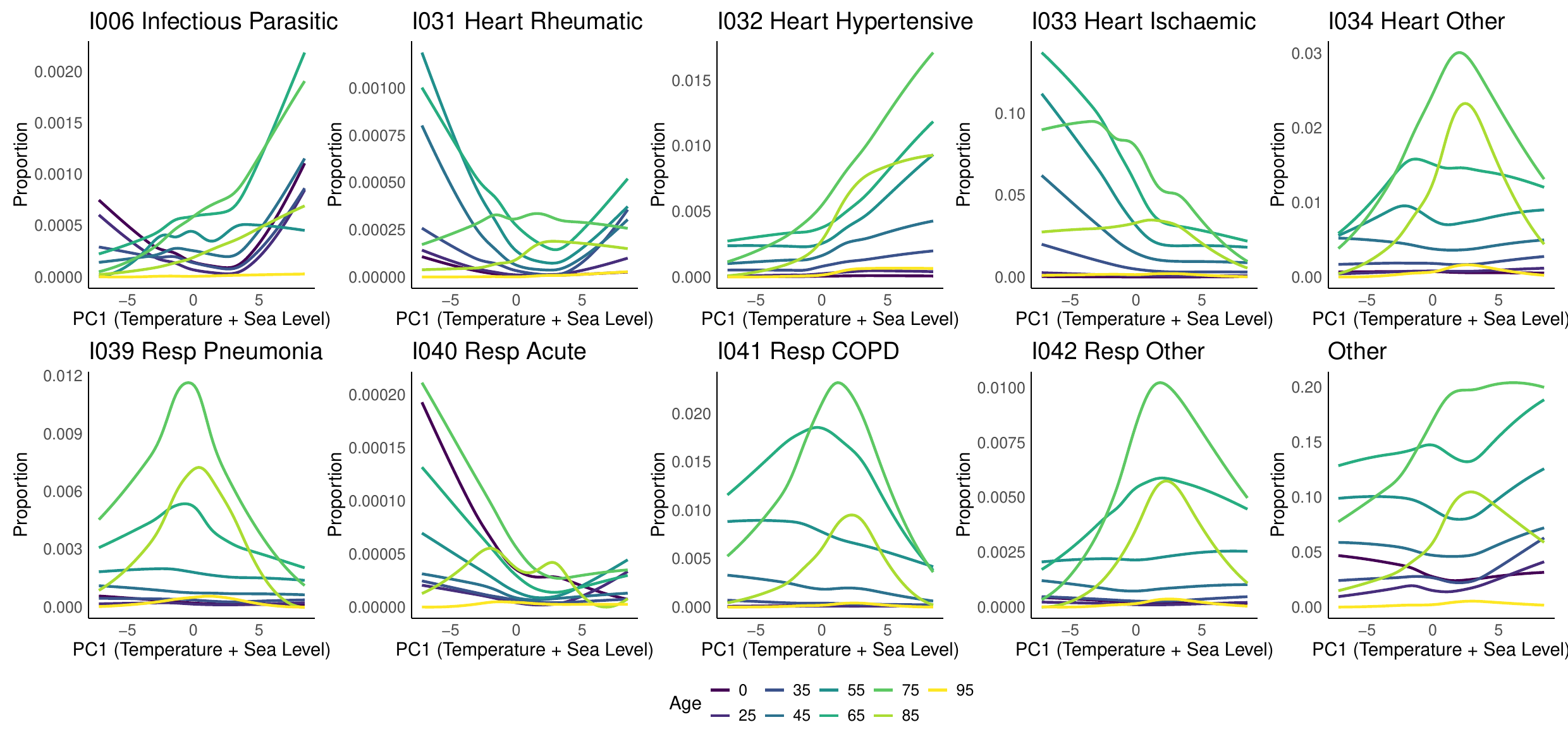}
\caption{Scenario analysis for US males, using the fitted CODA-GAM and based on varying PC1 from -8.0 to 8.0.}\label{fig:US_M_PC1_Sensitivity}
\end{figure}

Figures~\ref{fig:US_F_PC1_Sensitivity} and~\ref{fig:US_M_PC1_Sensitivity} present the results of the above scenario analysis in terms of the proportion of deaths for each cause and age for females and males in the US population, respectively, where each curve represents an age band within the model with the most notable differences by age within each cause. 

The proportions tend to increase with PC1 for Hypertensive and infectious parasitic (which includes vector-borne) diseases. Trends for some causes (COPD, pneumonia, and other heart and respiratory diseases) suggest that the proportion of deaths increases at lower levels of PC1 and then decreases as PC1 increases further. Other causes also exhibit this decreasing behaviour, namely Ischaemic heart and Rheumatic heart disease, and acute respiratory infection. This may be driven by the compositional nature of the data, meaning increases in Hypertensive and vector-borne diseases are likely being offset by decreases in other causes as PC1 increases. 

\begin{figure}[!htb]
\centering
\begin{subfigure}[b]{\textwidth}
\centering
\includegraphics[width=16cm]
{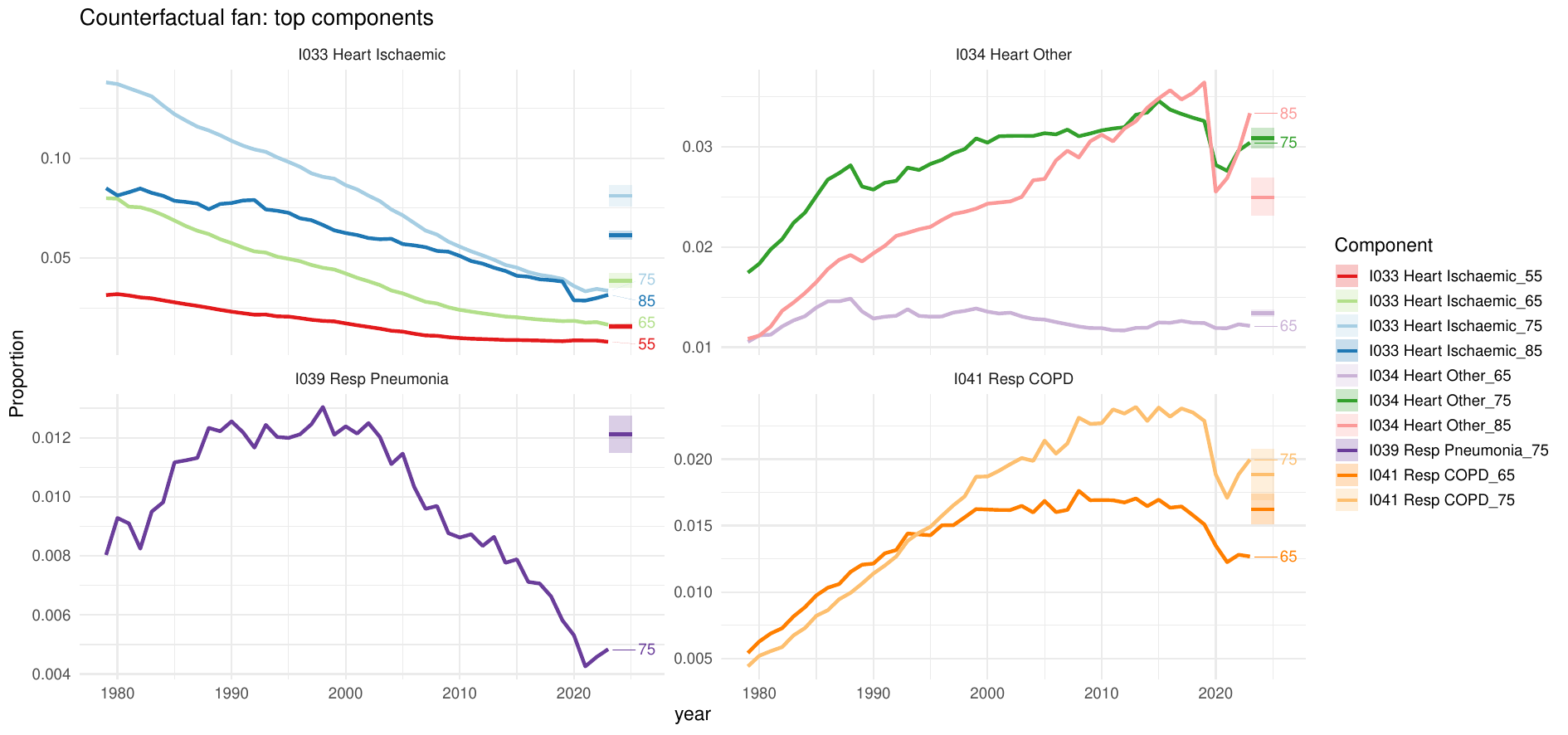}
\caption{Sensitivity analysis for US females, using the CODA-GAM fit, for the top ten combinations of cause and age bands. The horizontal line with the shaded region at the end of the panel  represents the expected proportion of deaths for a value of PC1 that is around twice the historic mean}\label{fig:US_F_Fan}
\end{subfigure}
\vspace{1em}
\begin{subfigure}[b]{\textwidth}
\centering
\includegraphics[width=16cm]
{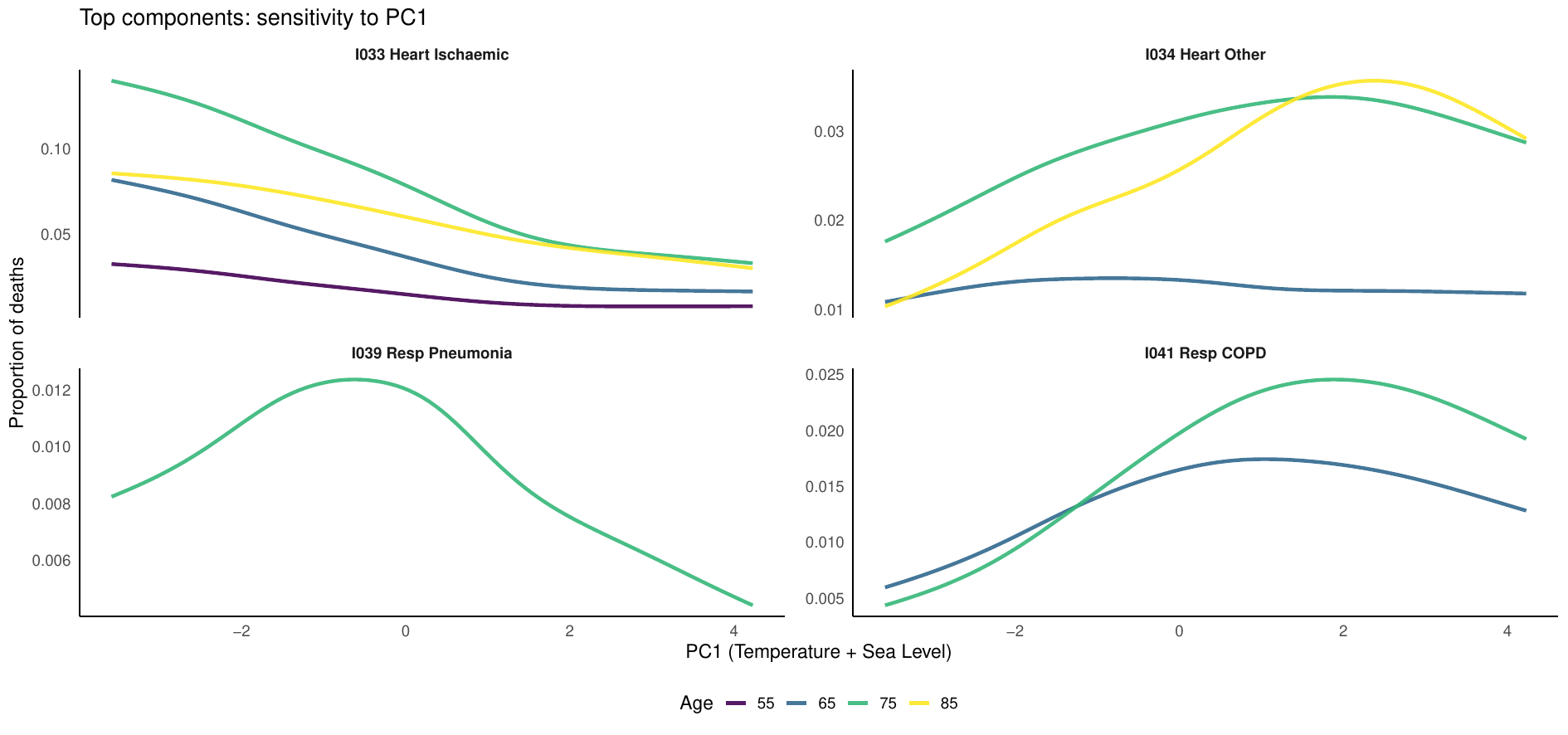}
\caption{Sensitivity of PC1 on proportion of deaths for top ten causes and age bands for females}\label{fig:US_F_Range}
\end{subfigure}
\caption{Sensitivity analysis for US females, using the CODA-GAM fit, for the top ten combinations of cause and age bands. }
\label{fig:US_F_Counterfactual}
\end{figure}

\begin{figure}[!htb]
\centering
\begin{subfigure}[b]{\textwidth}
\centering
\includegraphics[width=16cm]
{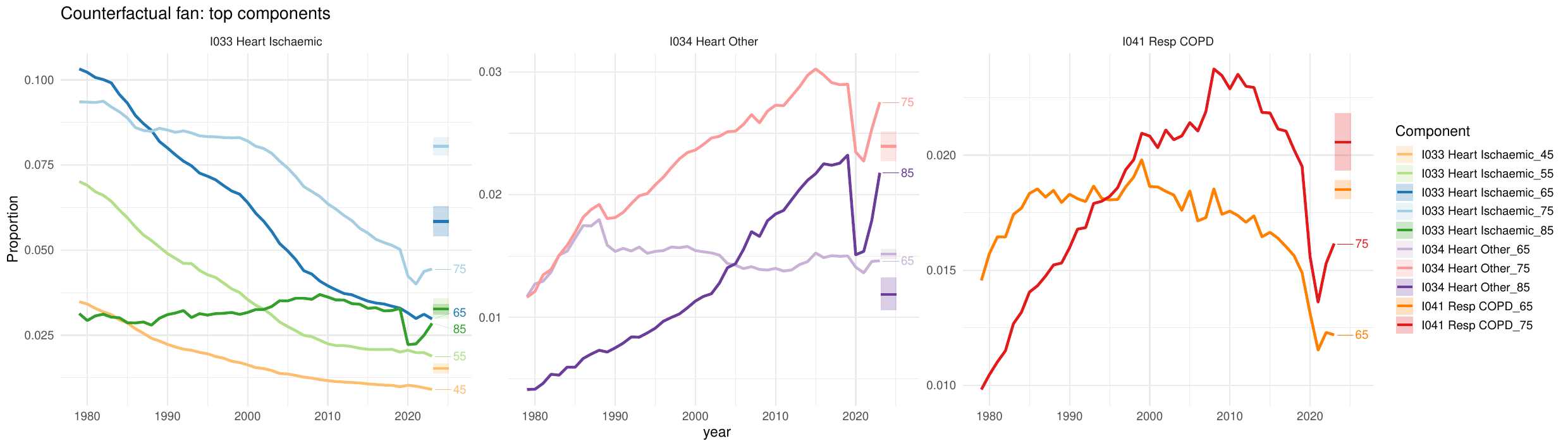}
\caption{Sensitivity analysis for US males, using the CODA-GAM fit, for the top ten combinations of cause and age bands. The horizontal line with the shaded region at the end of the panel  represents the expected proportion of deaths for a value of PC1 that is around twice the historic mean.}\label{fig:US_M_Fan}
\end{subfigure}
\vspace{1em}
\begin{subfigure}[b]{\textwidth}
\centering
\includegraphics[width=16cm]
{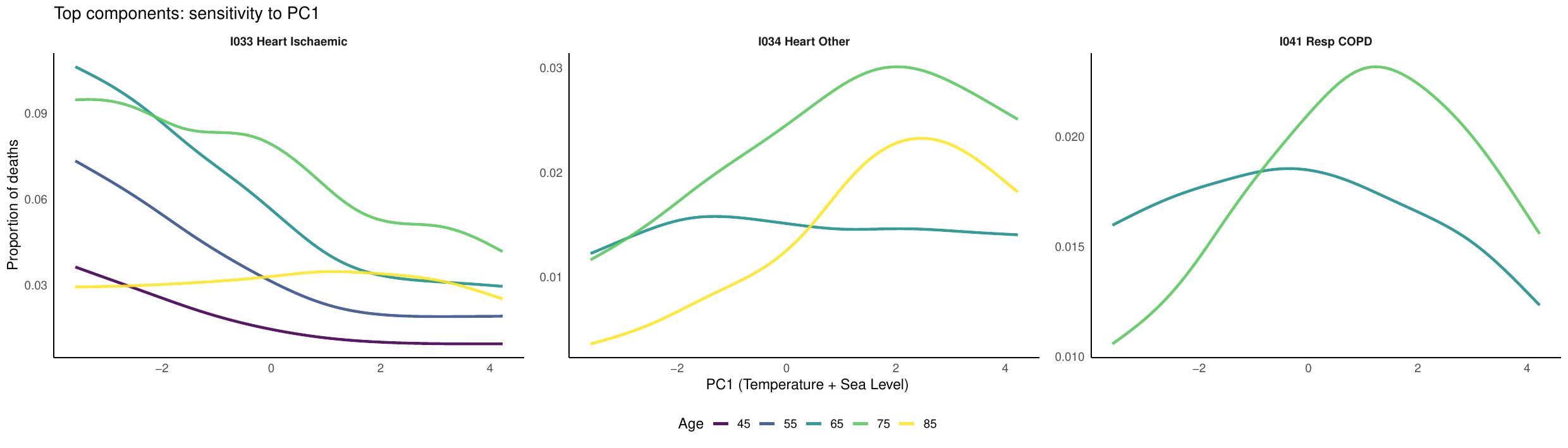}
\caption{Sensitivity of PC1 on proportion of deaths for top ten causes and age bands for males}\label{fig:US_M_Range}
\end{subfigure}
\caption{Sensitivity analysis for US females, using the CODA-GAM fit, for the top ten combinations of cause and age bands. }
\label{fig:US_M_Counterfactual}
\end{figure}

Figures~\ref{fig:US_F_Fan} and~\ref{fig:US_M_Fan} present the historical time series by cause and age, where each panel is a cause, and each line within the panel represents an age band. Note the groups are selected based on the top ten cause and age group combinations, as per Figure~\ref{fig:compositional_data}, meaning there are four causes for females (Ischaemic heart disease, Other heart disease, COPD, and Pneumonia), and three causes for males (Ischaemic heart disease, Other heart disease, and COPD). The horizontal line with the shaded region at the end of the panel  represents the expected proportion of deaths for a value of PC1 that is twice the historic mean.
We illustrate this method of sensitivity analysis as an important application of our new approach, and a step towards climate scenario analysis. 
Meanwhile, Figures~\ref{fig:US_F_Range} and~\ref{fig:US_M_Range} are a subset of Figures~\ref{fig:US_F_PC1_Sensitivity} and~\ref{fig:US_M_PC1_Sensitivity}, where this subset corresponds to each of the causes shown in Figures~\ref{fig:US_M_Fan} and~\ref{fig:US_F_Fan}, respectively, and only focuses on the historically observed range of PC1, i.e. approximately $-4.0$ to $+4.0$. 
Each line in the charts below represents the possible changes in the proportion of death for historically observed changes in PC1. It is clear that the historical proportions of deaths by cause and age closely follow the historical trends in PC1.

From this analysis, results are consistent with those shown earlier in Figure~\ref{fig:US_F_PC1_Sensitivity} and~\ref{fig:US_M_PC1_Sensitivity}. For females, small increases in PC1 to 2 times the historic mean suggests increases in the proportion of deaths across most causes and ages in the top 10, with exceptions being COPD for ages 75 - 84, and other heart diseases for ages 85 - 94. Similarly for males, modest increases in PC1 to twice the historical mean suggest increases in the proportion of most causes and ages in the top 10, with the exceptions being other heart diseases for ages 75 - 94. As values of PC1 varies from -4.0 to +4.0, similar behaviours we already start to see similar patterns as the analysis shown in Figures~\ref{fig:US_F_PC1_Sensitivity} and~\ref{fig:US_M_PC1_Sensitivity}. For example, for COPD, pneumonia, other respiratory and heart causes, increases in PC1 drive decreases in the proportion of deaths at older ages, and drive increases in the proportion of deaths at younger ages. Results reinforce varying effects by age, and the lower proportion of deaths for ages over 85 suggests possible acceleration of deaths. As PC1 increases, driven by high temperature extremes and sea level increases, the proportion of deaths for these causes decrease, and are offset in other causes. This approach to scenario analysis thus provides useful insight into the potentially offsetting impacts across the exposure population.


\section{Conclusion}\label{sec:discussion}

In light of the unprecedented uncertainty arising from climate-related risks, it is increasingly difficult for life insurers to rely on traditional methods of analysis by extrapolating general trends from historical data. Scenario and sensitivity analysis is becoming an increasingly important method to understand the impact of climate-related risks for banks and insurers, including life insurers. One further benefit of scenario analysis is that it enables the analysis of known relationships between climate-related risk factors and responses, as well as the variability of these relationships.

In this article, we adopt CODA techniques to understand the impacts from climate-related factors on the composition of deaths by cause and age bands, and subsequently perform scenario analysis for US death counts by cause from 1979 to 2023. Within the CODA construct, we applied the $\alpha$-transformation to remove the compositional constraint in a way that can handle zero counts in a principled manner. Second, to define climate-related covariates, we adopted PCA as a way to handle collinearity and offer a means of dimension reduction. We coupled CODA with a GAM approach to better reflect the non-linear relationships between climate factor-driven principal components and the composition of deaths.


Results from CODA-GAM suggests a combination of linear and non-linear effects of climate, which differ by cause and age for the climate and economic principal components. Notable results include the finding that Ischaemic heart disease tends to be positively impacted by temperature and sea level/dry extremes, whereas other respiratory causes show some possible offsetting impact. The results produced using US data also suggest that climate factors have more pronounced impacts that differ by age and cause, while macroeconomic factors tend to have fewer offsetting impacts between ages and causes. In the GAM models, older ages tend to have improved fit, likely because they have more observations in each cause- and age-band bucket. From a life insurer perspective, further analysis between the offsetting impacts could inform the implications around the natural hedge between a protection and annuities portfolio and inform the risk implications based on portfolio exposures by age and climate exposure.


Using the fitted CODA-GAM, we performed sensitivity analysis using a counterfactual range for the first principal component. The scenario reflects a range of PC1, where an increase in PC1 represents an increase in heat extremes, a decrease in cold extremes, and an increase in rainfall. Results of our scenario testing indicate that for the top ten climate-related cause-age combinations, Ischaemic heart disease is expected to show an increase in the proportion of deaths. There are expected offsetting impacts for the COPD cause in females, and the Other heart disease cause in males. 

There are several ways in which this research can be further extended, and we briefly mention a few below. First, in the context of climate scenario analysis, the approach explored in this paper can be further refined to calibrate more bespoke scenarios through the change in principal components, which allows climate factors to interact and vary simultaneously. This could also include applying the same method to other principal components identified. That is, while we performed sensitivity analysis using only PC1, further exploration into the results from PC2 and PC3 could enrich the insights and analysis from scenario modelling.

As discussed in Section~\ref{sec:model_results}, results indicate that there are non-linear effects, with some cause and age groups exhibiting a U-shape for increasing values of PC1. Further investigation into the medical or climate science analysis could shed more light on the decrease in the proportion of deaths for particular values of PC1. 
Further subgroup analysis of specific age bands and causes is also of interest, with initial results suggesting there is likely value in considering the composition of age and causes in climate scenario modelling. For example, moderately high values of PC1 (up to two standard deviations from the historical mean) lead to higher proportions of deaths for COPD for males and females aged 75-84; however, the same levels of PC1 can lead to lower proportions of deaths for Pneumonia. Conversely, causes such as Hypertensive and infectious parasitic diseases tend to increase with increasing PC1 across the majority of age bands (most notably for ages between 65 - 84), whereas causes such as acute respiratory and Ischaemic heart disease show signs of decreasing as PC1 increases. The gradients, i.e. speed of increase and decrease as PC1 increases, differ by age band. One further consideration is the extent to which such increases of deaths in the ages 65 - 84 age bands reflect an acceleration of deaths in these ages, and therefore fewer deaths in the oldest age bands.
More broadly, further research into the interaction between temperature extremes and cardiovascular causes, and potentially specific causal links to climate factors, could help explain the trends that appear by cause and age, and the peaks and troughs shown in Figures~\ref{fig:US_M_PC1_Sensitivity} and~\ref{fig:US_F_PC1_Sensitivity}. This may include a wide range of professionals: medical professionals, climate scientists, and actuaries.


Finally, an important methodological area for further refinement is the adjustment of the $\alpha$ parameter used in the transformation. For this analysis, we have selected $\alpha = 0.5$ for a balance between fit and forecast accuracy; however, $\alpha$ can be further tuned and refined depending on the ultimate goal \citep[i.e., minimising forecast errors through cross validation,][]{DSH+25}. Alternative approaches to CODA-GAM that could also be explored and considered include Dirichlet Regression and adopting machine learning techniques to further enhance flexibility in capturing non-linear relationships in the $\alpha$-space.

\section*{Acknowledgments}

The authors are grateful for the comments of the participants at the 1\textsuperscript{st} ASTIN Bulletin Conference at ETH Zurich in 2026, the questions and feedback from the Workshop on Insurance Mathematics at the University of Toronto in 2026, and thank the participants at a departmental seminar and a research day workshop within the Centre for  Emerging Risk at Macquarie University for their comments and questions. 
FKCH is supported by an Australian Research Council Discovery Project (DP240100143). HLS is supported by an Australian Research Council Future Fellowship (FT240100338).

\section*{Data Availability Statement}

Data and code supporting the findings of this study are publicly available at \url{https://github.com/zm-dong/coda_cause_mortality}.

\newpage
\appendix
\section{Explanatory variables} \label{app:EDA}

\begin{figure}[!htb]
\centering
\includegraphics[width=9cm]
{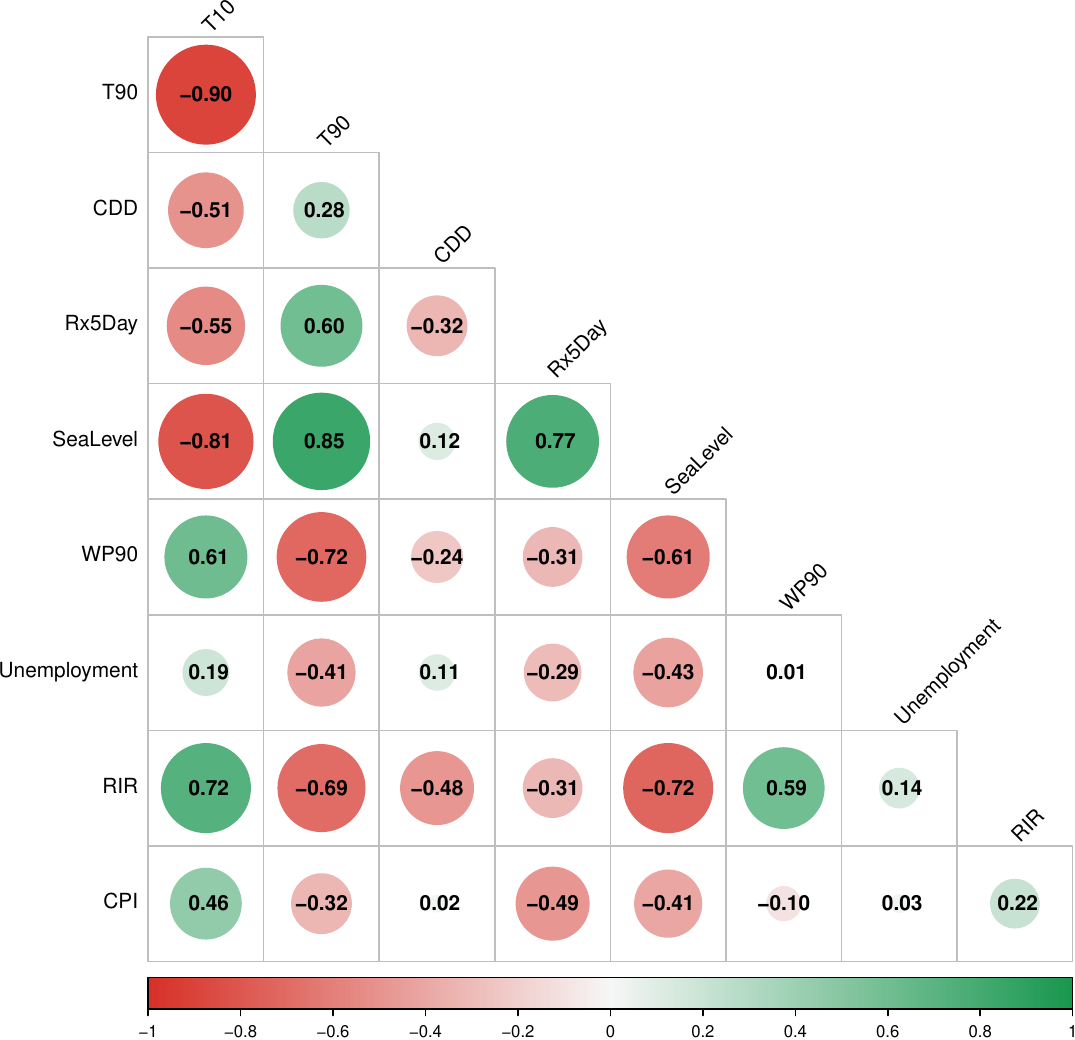}
\caption{Correlation matrix of explanatory variables for the US data. The size and colour of the circle represent the strength of the correlation.} \label{fig:correlation-matrix}
\end{figure}

Figure~\ref{fig:correlation-matrix} visualises the correlations between the underlying variables for the US data, before performing PCA. Table~\ref{tab:importance_pc} summarises the standard deviation, proportion of total variance explained, and cumulative proportion of total variance explained for the principal components.

\begin{table}[!htb]
\centering
\resizebox{\textwidth}{!}{
\begin{tabular}{@{}lccccccccc@{}}
\toprule
\textbf{Importance of components} & \textbf{PC1} & \textbf{PC2} & \textbf{PC3} & \textbf{PC4} & \textbf{PC5} & \textbf{PC6} & \textbf{PC7} & \textbf{PC8} & \textbf{PC9} \\
\midrule
\textbf{SD} & 2.1574 & 1.3010 & 1.0355 & 0.9233 & 0.5820 & 0.4825 & 0.3069 & 0.1970 & 0.1530 \\
\textbf{Proportion of Variance} & 0.5172 & 0.1881 & 0.1192 & 0.0947 & 0.0376 & 0.0259 & 0.0105 & 0.0043 & 0.0026 \\
\textbf{Cumulative Proportion} & 0.5172 & 0.7052 & 0.8244 & 0.9191 & 0.9568 & 0.9826 & 0.9931 & 0.9974 & 1.0000 \\
\bottomrule
\end{tabular}}
\caption{Importance of Principal Components}
\label{tab:importance_pc}
\end{table}

The principal component analysis indicates that the first component explains 51.72\% of the total variance, representing a dominant underlying factor in the data. PC1 here reflects the temperature and sea level extremes in the underlying data. The second and third components account for an additional 18.81\% and 11.92\%, respectively, bringing the cumulative explained variance to 82.44\% for the first three components. The second component is associated with rainfall variability, particularly the contrast between cumulative dry days and extreme precipitation, whereas the third component reflects macroeconomic variability through inflation and unemployment.

\newpage
\section{CODA-LM}
\label{sec:coda-lm}

The results of the CODA-LM summarised in Table~\ref{fig:LMCODA_US_Results_Commentary} suggest that older ages tend to be more impacted by climate-related regression factors. Regression coefficients suggest that, overall, the impacts on US males are similar to those on US females, having more pronounced increases overall for hypertensive, COPD, and Ischaemic heart disease causes, but less pronounced differences by age within these causes. 

Further considerations include scenario analysis using CODA-LM, similar to the application in Section~\ref{sec:scenarios}.

\begin{small}
\begin{longtable}{|p{4cm}|p{6cm}|p{5cm}|}
\hline
\textbf{Factor} & \textbf{Female Results} & \textbf{Male Results}\\
\hline\hline
\endfirsthead
\hline
\textbf{Factor} & \textbf{Female Results} & \textbf{Male Results}\\
\hline\hline
\endhead
\hline
PC1 (Temperature and sea level) & This factor increases the proportion of COPD, other respiratory, and hypertensive deaths. There is a small impact on infectious parasitic causes. Older ages tend to be more impacted, and the increase in older age deaths for these causes is offset by younger age deaths. & Similar impacts to US females, but the increases for COPD and other respiratory deaths are even more concentrated at older ages. \\
\hline
PC2 (Dry and Wet) & This factor drives a small increase for older age bands for respiratory acute causes of death. & Similar to US females. \\
\hline
PC3 (Macroeconomic and Wind) & This factor drives a decrease in older age deaths, offset by an increase in younger age deaths for respiratory acute causes of death and heart rheumatic causes of death. & Similar to US females, but the age differences are less pronounced. There is a small increasing impact on hypertensive deaths.\\
\hline
\caption{CODA-LM regression results for US female and male data using the $\alpha$-transformation}\label{fig:LMCODA_US_Results_Commentary}
\end{longtable}
\end{small}

\newpage
\section{US GAM \texorpdfstring{$R^2$}{R2} and goodness of fit}\label{sec:Appendix_US_GAM_Fit}

\begin{figure}[!htb]
\centering
\begin{subfigure}[b]{\textwidth}
\centering
\includegraphics[width=14.6cm]
{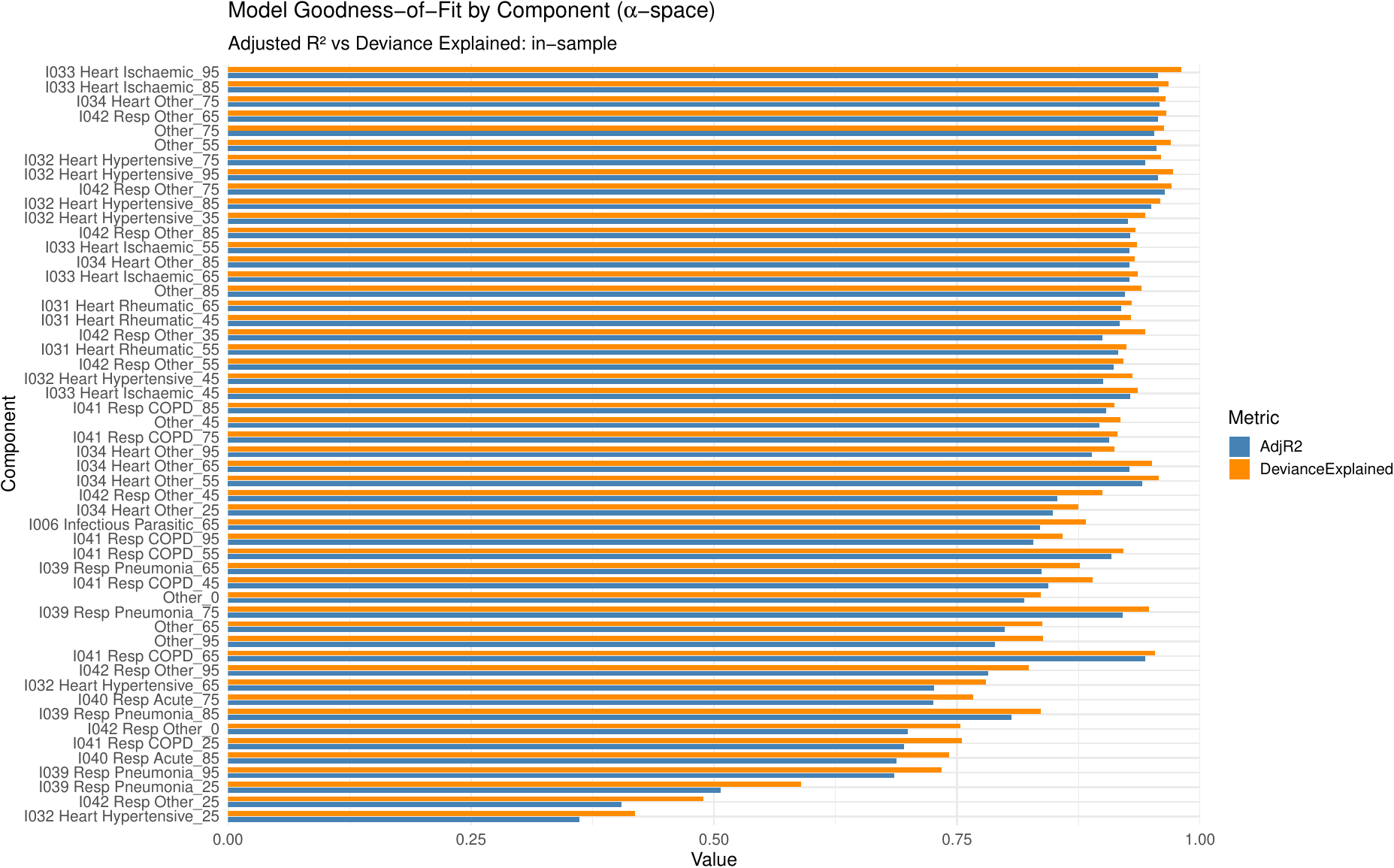}
\caption{GAM R-squared, and percentage deviation explained for US females in the $\alpha$-space.}
\end{subfigure}
\vspace{1em}
\begin{subfigure}[b]{\textwidth}
\centering
\includegraphics[width=14.6cm]
{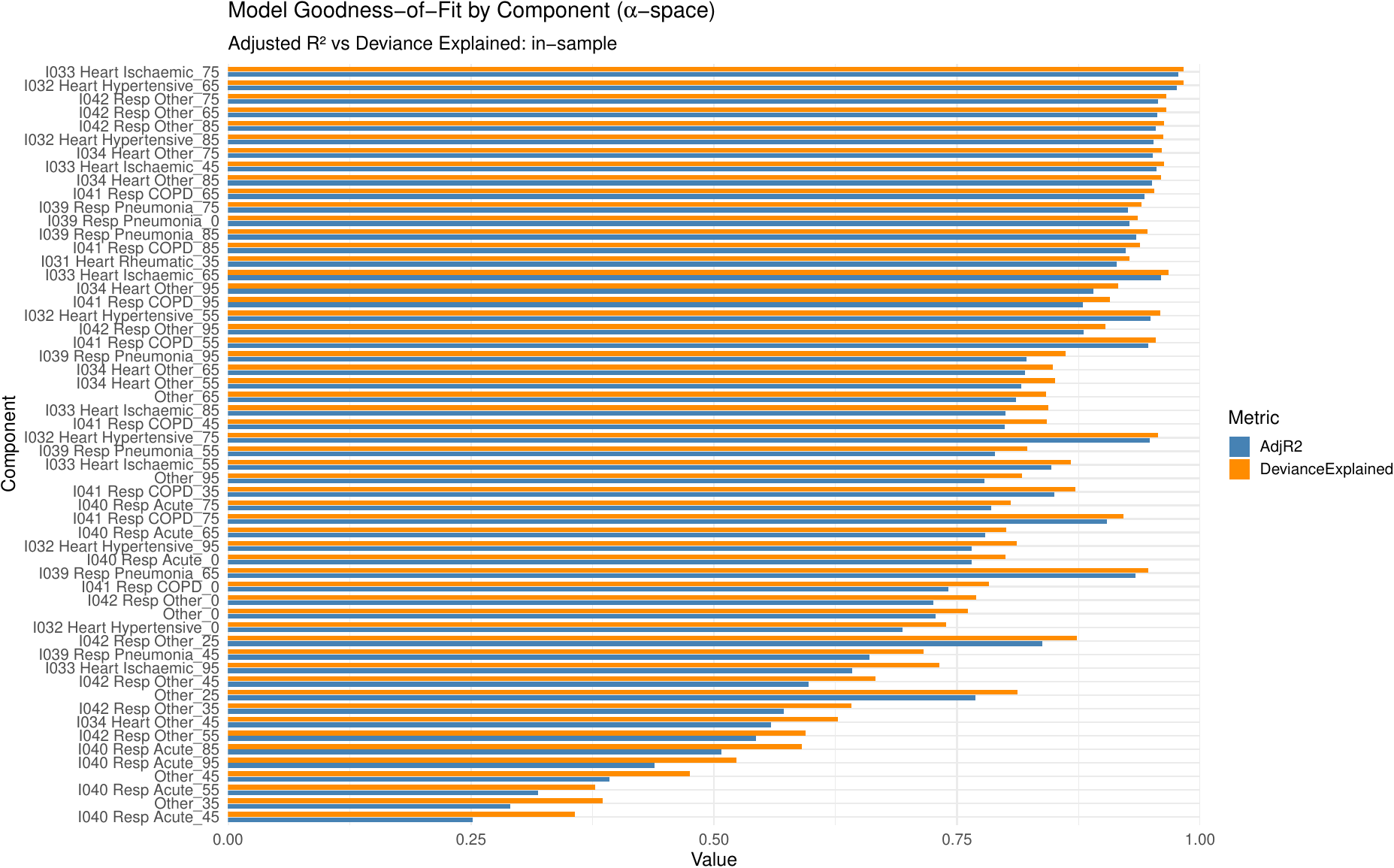}
\caption{GAM R-square,d and percentage deviation explained for US males in the $\alpha$-space.}
\end{subfigure}
\caption{GAM R-squared and percentage deviation explained in the $\alpha$-space.}
\label{fig:US_GAM_R2}
\end{figure}

Figure~\ref{fig:US_GAM_R2} shows the $\alpha$-space $R^2$ for US females and males. As this $R^2$ is produced in the $\alpha$-space, the cause and age labels are determined by mapping back to the correlations between the $\alpha$-transformed axis and the original component.

\begin{figure}[!htb]
\centering
\begin{subfigure}[b]{\textwidth}
\centering
\includegraphics[width=14.6cm]
{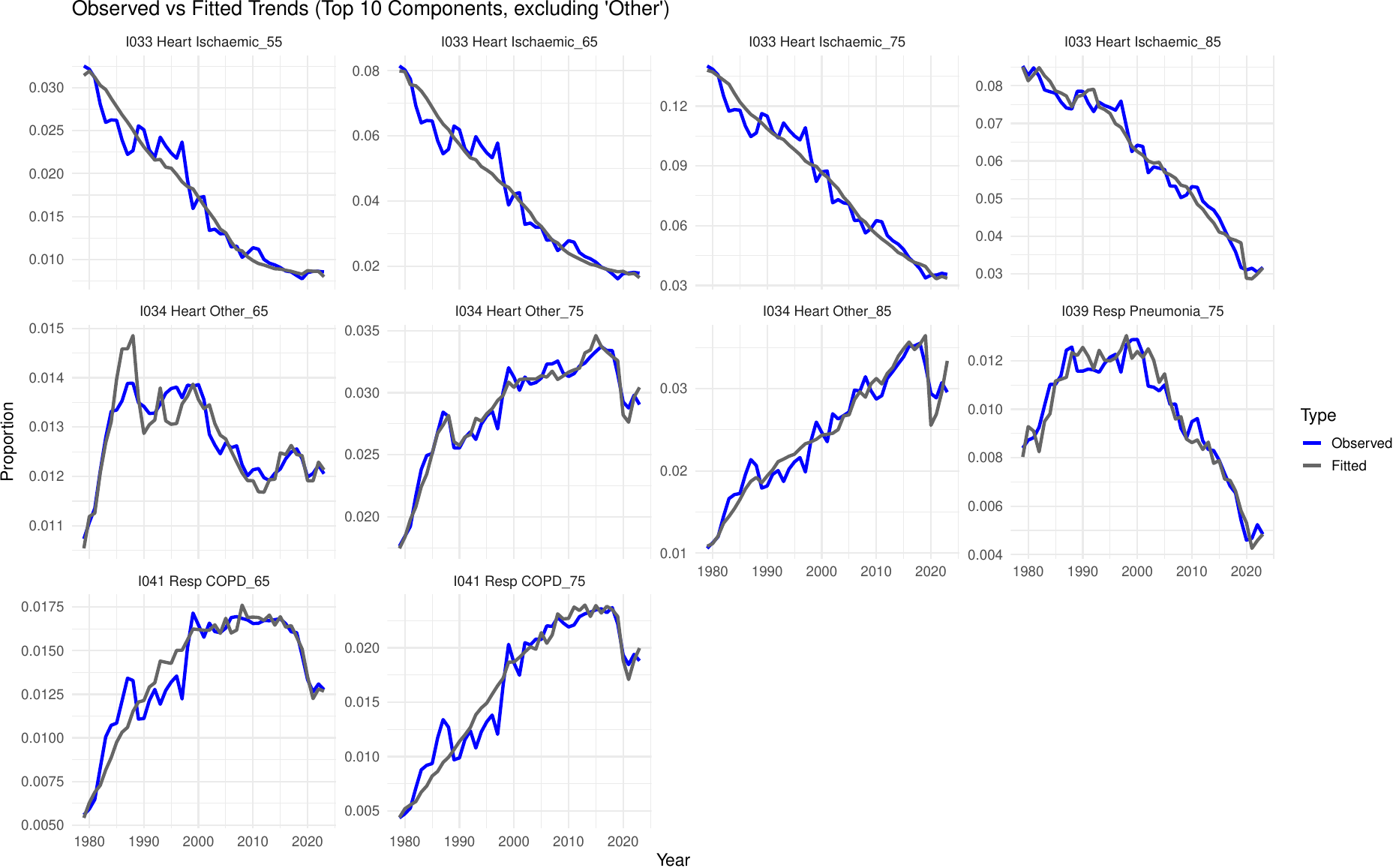}
\caption{GAM fit for US females, top ten age and cause combinations. The blue lines show the observed data, and the grey lines indicate the model fit.}
\end{subfigure}
\vspace{1em}
\begin{subfigure}[b]{\textwidth}
\centering
\includegraphics[width=14.6cm]
{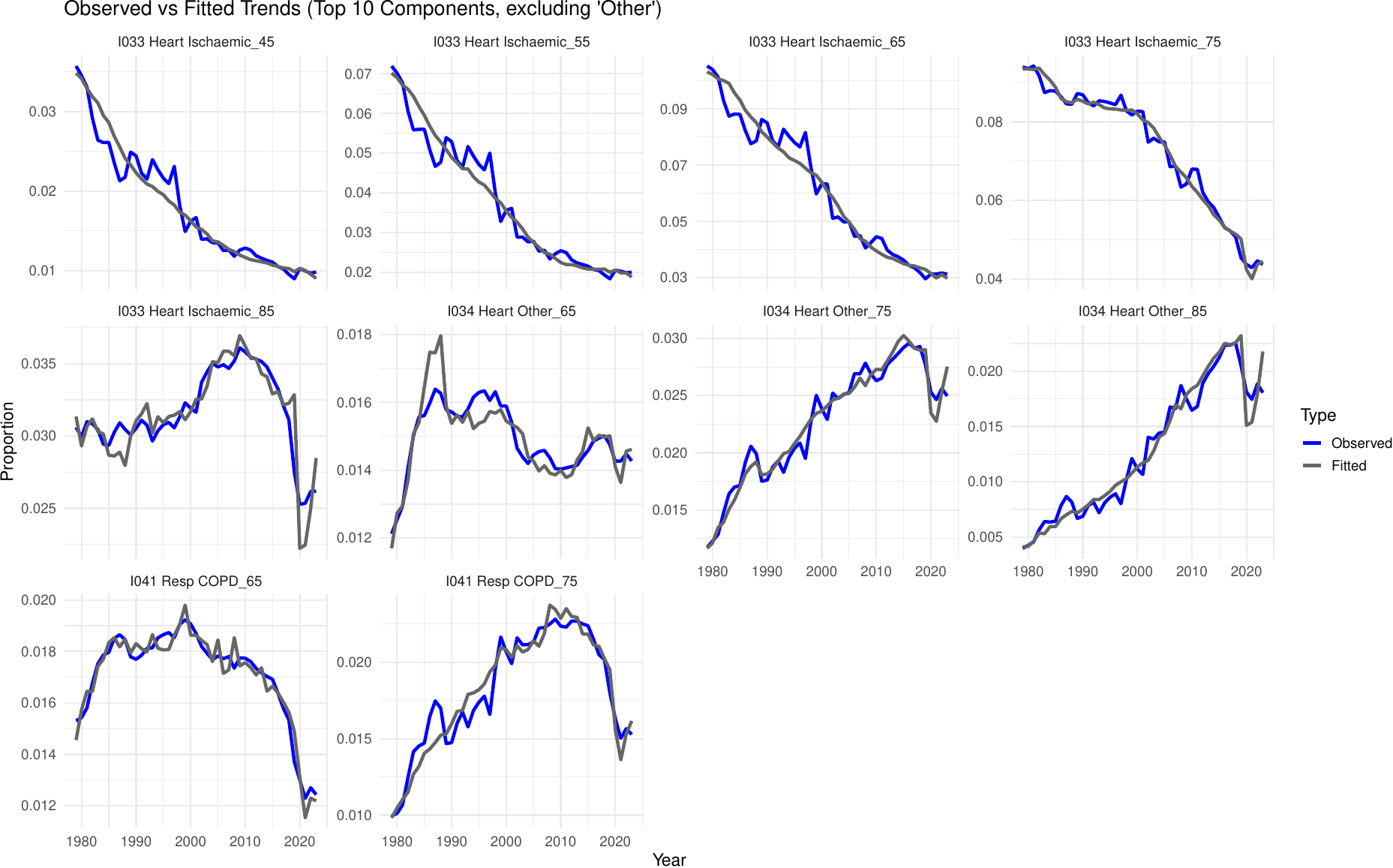}
\caption{GAM fit for US males, top ten age and cause combinations. The blue lines show the observed data, and the grey lines indicate the model fit.}
\end{subfigure}
\caption{GAM fit for US females and males, top ten causes and age bands.}
\label{fig:US_GAM_Fit}
\end{figure}

The plots of the actual against fitted GAM for the top ten cause and age bands are shown in Figure~\ref{fig:US_GAM_Fit} for the female and male data in the US to illustrate the fit in the top cause groups. Similarly, this fit is shown in the $\alpha$-space, rather than the back-transformed Aitchison space.

\newpage
\bibliographystyle{apalike}
\bibliography{CoDA.bib}

\end{document}